%% file: estimating_under5_mortality.tex
\documentclass[12pt]{article}
\usepackage[utf8]{inputenc}

\RequirePackage{calc}
\RequirePackage{natbib}
\RequirePackage[american]{babel}
\usepackage{chngcntr}

\usepackage{setspace}
\usepackage{array}

\usepackage{rotating}
\usepackage{color}
\usepackage{url}
\RequirePackage{amsthm,multirow,amsmath,fancybox,amssymb}

\usepackage{graphics}

\usepackage{moreverb,url}

\newcommand\BibTeX{{\rmfamily B\kern-.05em \textsc{i\kern-.025em b}\kern-.08em
T\kern-.1667em\lower.7ex\hbox{E}\kern-.125emX}}
\usepackage{gensymb}
\include{definitions}

\usepackage{tensor}
\newcommand{\qf}{\tensor*[_5]{\mbox{q}}{_0}}

\title{\Large \textbf{Estimating Under Five Mortality in Space and Time} \\ \textbf{in a Developing World Context}}
\author{\vspace{-2mm}
Jon Wakefield$^{1,2}$, Geir-Arne Fuglstad$^{3}$, Andrea Riebler$^{3}$, Jessica Godwin$^{1}$, \\
Katie Wilson$^{2}$ and Samuel J. Clark$^{4}$\\
\vspace{-2mm}
{\small $^{1}$Department of Statistics,
  University of Washington, Seattle}\\
  \vspace{-2mm}
{\small $^{2}$Department of Biostatistics,
University of Washington, Seattle}\\
\vspace{-2mm}
{\small $^{3}$Department of Mathematical Sciences, Norwegian University of Science and Technology, Trondheim}\\
\vspace{-2mm}
{\small $^{4}$Department of Sociology,
The Ohio State University, Columbus}
}
\date{}

\begin{document}

\maketitle

\begin{abstract}
Accurate estimates of the  under-5 mortality rate (U5MR) in a developing world context are a key barometer of the health of a nation.
This paper describes new models to analyze survey data on mortality in this context. We are interested in both spatial and temporal description, that is, wishing to estimate U5MR across regions and years, and to investigate the association between  the U5MR and spatially-varying covariate surfaces. We illustrate the methodology by producing yearly estimates for subnational areas in Kenya over the period 1980--2014 using data from demographic health surveys (DHS). We use a binomial likelihood with fixed effects for the urban/rural stratification to account for the complex survey design. We carry out smoothing using Bayesian hierarchical models with continuous spatial and temporally discrete components. A key component of the model is an offset to adjust for bias due to the effects of HIV epidemics.  Substantively, there has been a sharp decline in U5MR in the period 1980--2014, but large variability in estimated subnational rates remains. A priority for future research is understanding this variability.
 Temperature, precipitation  and a measure of malaria infection prevalence were candidates for inclusion in the covariate model. \end{abstract}


\maketitle
\section{Introduction}\label{sec:introduction}

Currently UNICEF estimates the under-five child mortality rate (U5MR) at the national level (which is known as Admin 0), using the Bayesian B-spline bias-reduction (B3) method \citep{alkema2014child,alkema:new:14}. However, subnational variation is of great interest, and has been highlighted as such in the Sustainable Development Goals (SDGs). SDG 3.2 states, ``By 2030, end preventable deaths of newborns and children under 5 years of age, with all countries aiming to reduce neonatal mortality to at least as low as 12 per 1,000 live births and under-5 mortality to at least as low as 25 per 1,000 live births.''
From \url{https://sustainabledevelopment.un.org/post2015/transformingourworld}, with reference to review processes, paragraph 74.g states, ``They will be rigorous and based on evidence, informed by country-led evaluations and data which is high-quality, accessible, timely, reliable and disaggregated by income, sex, age, race, ethnicity, migration status, disability and geographic location and other characteristics relevant in national contexts."

In much of the developing world,  there is no vital registration, and estimates of U5MR are based on survey data. In this paper, we carry out detailed analyses of such data from Kenya. Intervention and subnational policy are generally implemented at Admin 2, the second administrative level. This naming convention can be confusing with the 8 provinces of Kenya not appearing in the Admin hierarchy, and the 47 counties being labeled Admin 1. It is at this level that policies are implemented for Kenya, and hence is our spatial target of inference.
We use data from Demographic Health Surveys (DHS). The 
DHS Program began in 1984 and has carried out more 300 surveys in over 90 countries. Typically stratified cluster sampling is carried out  information is collected on population, health, HIV and nutrition.

 We briefly review previous approaches to producing sub-national U5MR estimates. Adopting demographic notation, $_n q_x = \Pr( \mbox{ death in }[x,x+n) ~| ~\mbox{survival to }x )$, so that we are interested in $_5q_0$ (note that strictly speaking U5MR is a probability rather than a rate).
\cite{dwyer:etal:2014} compare various spatial models for U5MR modeling in Zambia using DHS data. The logit of the U5MR is modeled as normally distributed, but with a single common variance across all studies, which is clearly inappropriate since it does not acknowledge the differing effective sample sizes in each area. Computation was carried out using the integrated nested Laplace approximation (INLA) of \cite{rue:etal:09}.
\cite{mercer:etal:15} analyzed DHS data from 22 regions in Tanzania and assumed a likelihood in which the logit of the weighted (design) estimator was assumed to be normally distributed with variance given by the design variance. A discrete space, discrete time (5-year intervals) model \citep{knorrheld:00} was used to smooth the mean  of this distribution, with implementation via INLA. Unfortunately in our Kenya study we require  finer temporal and spatial scales, and at such scales the data are sparse and the weighted estimators are unstable, making the method of \cite{mercer:etal:15} unfeasible. 
 \cite{pezzulo:etal:17} model $\tensor*[_4]{\mbox{q}}{_1}$ across 27 countries in Sub-Saharan Africa, at the Admin 1 level. Estimation was based on the most recent DHS with the log weighted estimators assumed to be normally distributed with spatial smoothing being carried out via the model of \cite{leroux:etal:99}.
 Extensive covariate modeling was carried out with potential variables being averaged within areas, and also allowing interactions by large regions (with three regions in total). The associations at the area-level cannot be transferred to the individual-level as this opens up the possibility of the ecological fallacy \citep{wakefield:08}.

\cite{burke:etal:16} follow a different approach to modeling U5MR across sub-Saharan Africa.  
Kernel density estimation (KDE) is carried out with surfaces produced at a geographical scale of  approximately 10km$\times$10km. This approach follows \cite{larmarange:bendaud:14} who used the same method  in the context of HIV prevalence estimation. Inference, including producing uncertainty surfaces, is difficult to obtain with KDE and the 
approach has been   found to be inferior to Bayesian geostatistical modeling \citep{hallett:etal:16}.


More recently, \cite{golding:etal:17} carried out subnational estimation of U5MR for sub-Saharan Africa, with a continuous model for  space.
Four separate models were fitted to the age groups 0--1 months, 1--11 months, 12--35 months, 36--59 months, with the subsequent estimates being combined to give the U5MR. This combination is done by taking draws from the posteriors assuming they are independent, but they are not, since they are based on the same children. As well as  full birth history data from DHS, summary birth history is also included. These latter typically consist of the number of children ever born, and the number who have died, along with the age of the mother.  For full birth history the data are modeled as binomial with no explicit correction for the survey design. The summary birth histories are also assumed to be binomially distributed, with an artificial response and denominator created through an elaborate procedure.  A space-time smoothing model is specified via the stochastic partial differential equations (SPDEs) formulation of \cite{lindgren:etal:11}. The same space-time covariance parameters are assumed for the whole of Africa. Covariates are also modeled, and we give further details of the approach in Section \ref{sec:exploratory}. There is no adjustment for mothers lost to HIV, which can lead to serious underestimation in countries with HIV epidemics. Estimates in each spatial grid cell are adjusted so that the national total agrees with the Global Burden of Disease (GBD) estimates. The most recent GBD \citep{GBD:mortality:17} produced national estimates for 195 countries and territories over the period 1970--2016.
Some of the constituent data in the study of  \cite{golding:etal:17} do not contain GPS locations, but rather the administrative region within which the clusters were sampled. In this case,  Golding et al.~(2017, Supplementary Materials, Section 8) assign the data to a single point within the area, where this point is obtained as the weighted combination of $J$ representative points that are obtained through $k$-means clustering. This approach is, at best, an approximation, since one needs to take a mixture over the likelihoods at each potential location, see \cite{wilson:wakefield:17}.

The rest of this paper is structured as follows. In Section \ref{sec:data} we describe the data that we use for analysis. Section \ref{sec:constructing} develops the method and gives the results for constructing the space-time child mortality surface, while Section \ref{sec:exploratory} does the same for covariate modeling. Section \ref{sec:discussion} concludes the paper with a discussion of ways in which we would like to extend the model.


\section{Data}\label{sec:data}

\subsection{Survey Data}

\begin{figure}
\centering
\includegraphics[height = 10cm]{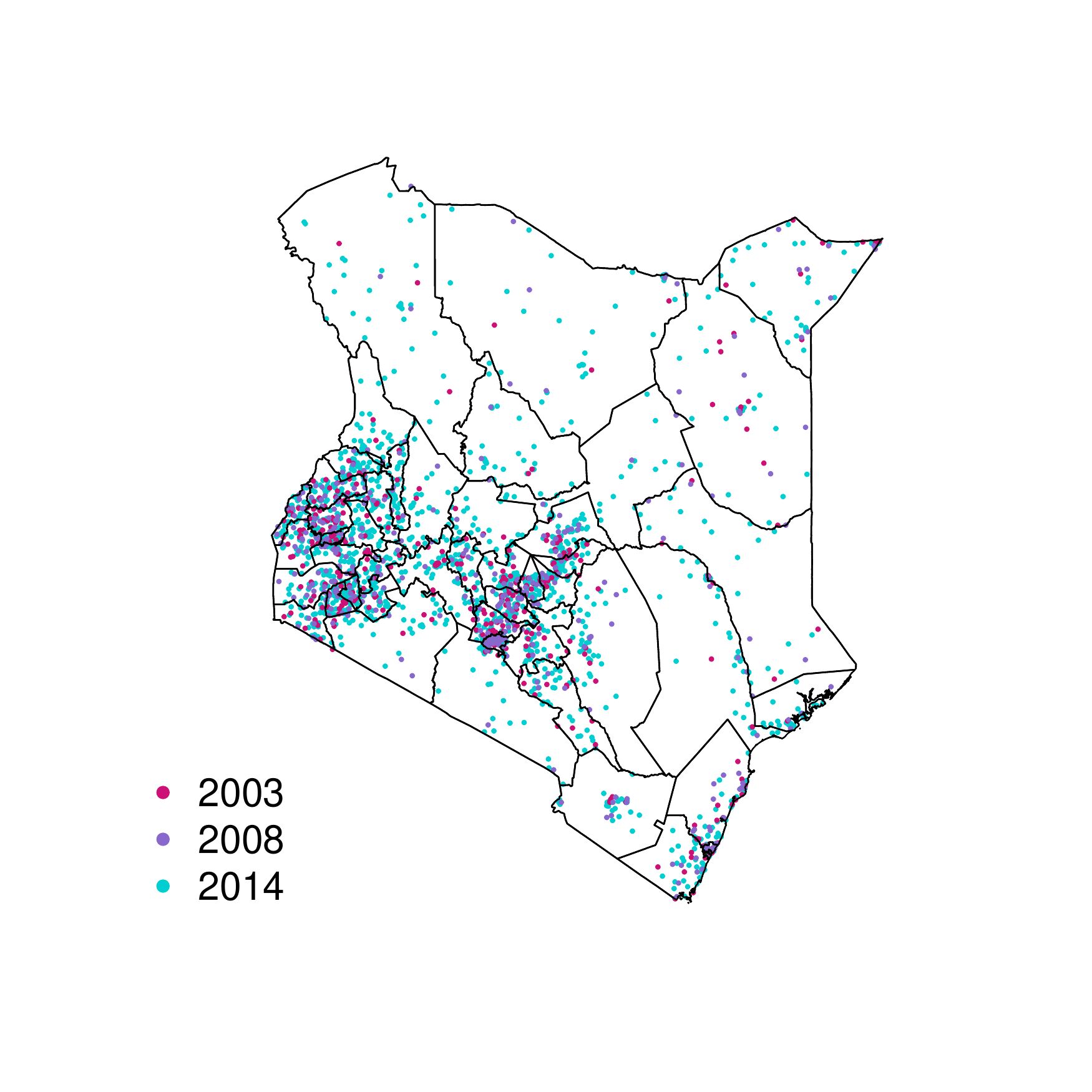}
\caption{Cluster locations in the three DHS that we consider, with boundaries of the 47 counties.}
\label{fig:ClusterLocs}
\end{figure}

To estimate child mortality in Kenya, we use data from three DHS conducted in 2003, 2008--2009 and 2014. Both the 2003 and 2008--2009 Kenya DHS were designed to give estimates  for the 8 provinces, and for urban and rural regions separately. To this end, the sample was stratified by 8 provinces crossed with an urban/rural designation to yield 15 strata (Nairobi is solely urban). In each of these surveys the first sampling stage selected 400 enumeration areas (EAs) from a sampling frame constructed from the 1999 Census. In the second stage for both the 2003 and 2008--2009 surveys, 10,000 households were selected within the sampled EAs. The 2014 Kenya DHS was designed to make estimates of demographic indicators at the 47 county level, so it was stratified by the 47 counties crossed with urban/rural indicators. This yields 92 strata since Nairobi and Mombasa are both entirely urban. The first sampling stage of the 2014 survey selects 1,584 EAs (that produced data that could be used) from the 92 strata using a sampling frame developed from the 2009 Census. In the second stage, 40,300 households were sampled from the selected EAs. 
To estimate U5MR we use the portion of the survey devoted to retrospective birth histories. Women who slept in the house the night before, and are aged 15--49 are asked to enumerate all births with dates of birth, and for children who have died, dates of death. Birth histories are converted into person months for each child in the dataset.  Using a discrete hazard model, each person month yields a Bernoulli (binary) random variable, survived/dead. Hence, we implement a discrete time event history analysis. It is important to note that each unique case can result in at most one death. 
Figure \ref{fig:ClusterLocs} shows the cluster locations for the three surveys along with the boundaries of the 47 counties. We see that the distribution of the sampling locations is far from uniform, reflecting population density.  Reported response rates for households and women are high. Such data are potentially subject to various biases, e.g.,~recall bias, as the birth histories may go back many years if the woman surveyed is old. Though we have data from only three survey waves, the retrospective birth history gives us data on births over the period 1980--2014. 
According to the final reports from the Kenya DHS, nationally the U5MR per 1,000 births was 115 in 2003, 74 in 2008--2009 and 52 in 2014. While these estimates show a clear trend of decreasing child mortality, we would like to investigate the subnational variability across the 47 counties.  Kenya provides a good test example due to  large number of clusters (1,584) sampled in the 2014 DHS. 




\subsection{HIV Adjustment}


Kenya has had a relatively high prevalence of HIV, and this can lead to serious bias in estimates of U5MR, particularly before antiretroviral therapy (ART) treatment became widely available. Pre-treatment HIV positive women had a high risk of dying, and such women who had given birth were therefore less likely to appear in surveys. The children of HIV positive women are also more likely to die before age 5 compared to those born to HIV negative women, and therefore we expect to underestimate U5MR if we do not adjust for the missing women, i.e.,~the missing data are non-ignorable.

Estimates of bias may be obtained using the cohort component projection model of 
 \cite{walker:etal:12}. Under this model, for a particular survey, year and province, the number of births is estimated, and these are attributed to HIV-negative and HIV-positive women, using estimates of the number of women in need of services to prevent mother-to-child transmission. The children born are then further subdivided into those that will and those that will not become infected with HIV, and survival probabilities of these children are then estimated, to produce a bias ratio.
Let $\qf_{l,k}(t)$ represent the  true U5MR and $\widetilde{\qf}_{l,k}(t)$ the biased (unadjusted for HIV) U5MR in  survey $k$, year $t$ and province $l$, $l=1,\dots,8$. The  \cite{walker:etal:12} method gives an estimate of,
\begin{equation}\label{eq:BIAS}
\mbox{BIAS}_{l,k}(t)= \frac{\qf_{l,k}(t)}{{\widetilde{\qf}}_{l,k}(t)} \geq 1.
\end{equation}

\begin{figure}[h]
\centering
\includegraphics[scale = 0.4]{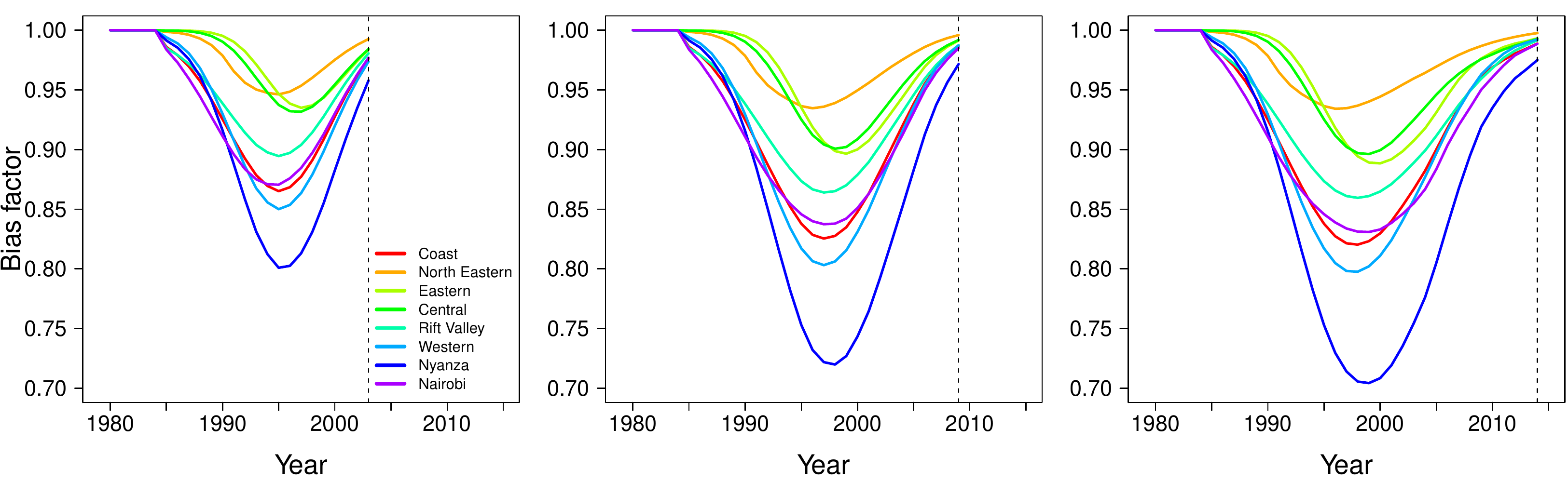}
\caption{HIV adjustment ratios of reported U5MRs to ``true" U5MRs, that is the reciprocal of (\ref{eq:BIAS}), by survey over time (left is 2003, middle is 2008--2009, right is 2014), and in eight regions. Ratios were obtained using the method of \cite{walker:etal:12}.}
\label{fig:hivadjust} 
\end{figure}

\begin{figure}[h]
\centering
\includegraphics[scale = 0.2]{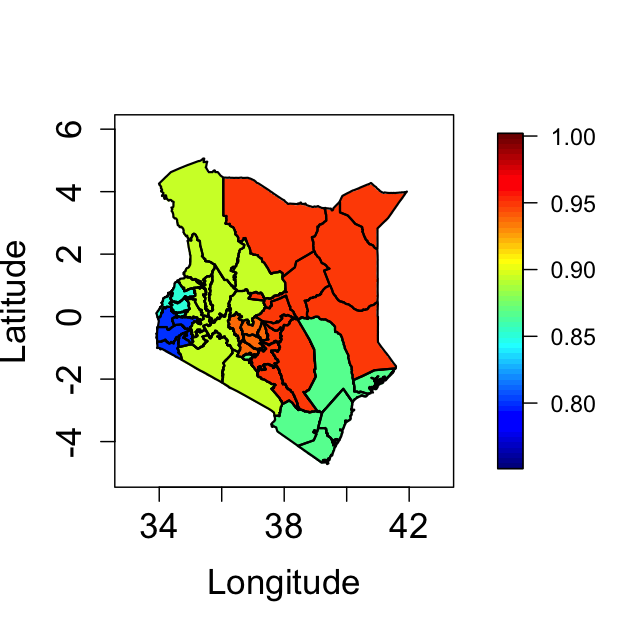}
\includegraphics[scale = 0.2]{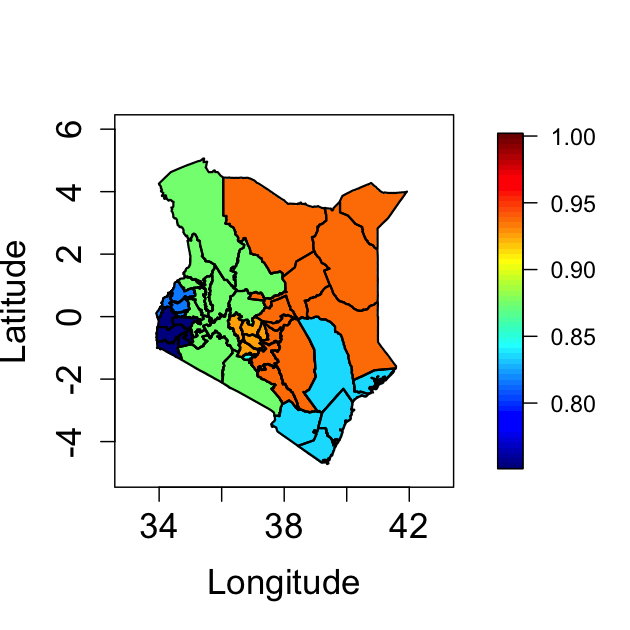}
\includegraphics[scale = 0.2]{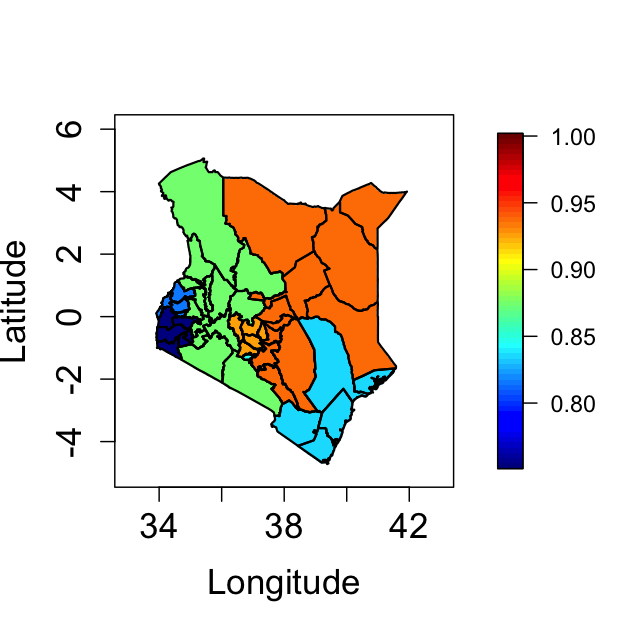}
\caption{Maps of HIV adjustment ratios of reported U5MRs to ``true" U5MRs, that is the reciprocal of (\ref{eq:BIAS}), by survey in 1995. The 3 columns represent the adjustments from the 2003, 2008--2009, 2014 surveys. Ratios were obtained using the method of \cite{walker:etal:12}.}
\label{fig:spatialHIVadj} 
\end{figure}

 Figure~\ref{fig:hivadjust} shows the ratios (reciprocal bias) plotted against year for each of the three surveys, and for the 8 regions of Kenya for which we have available data; we would prefer to have estimates at the 47 county level, but the constituent data are not available, and the 47 counties are nested within the 8 provinces, which eases the application of the adjustment.  We see that the ratios of reported to true decrease as the HIV epidemic takes hold and then increase with the uptake of ART.  Figure \ref{fig:spatialHIVadj} shows maps of the ratios in 1995, and the large between-province differences are apparent.
The ratios will clearly make a significant impact on our estimates, and are included in an offset in the model we describe in Section \ref{sec:constructing}. A current weakness of our approach is that we do not account for the uncertainty in the manner by which the ratios were estimated.

\section{Constructing a Space-Time Surface}\label{sec:constructing}

\subsection{The Space-Time Model}\label{sec:reconstruct:model}

Survey data come from and describe a finite population. The DHS provides sampling weights for each individual that account for the selection probability and non-response. \cite{skinner:wakefield:17} review the design and analysis of survey data. The design-based (or randomization) approach to inference is to couch inference in the context of repeated sampling from the fixed finite population. The word {\it fixed} is key here, the data are not viewed as random, rather the indices of the units (households, here) within the population that are sampled are the random variables. Weighted (often referred to as direct) estimators \citep{horvitz:thompson:52} provide a design-consistent approach to estimation, but the sparsity of data in both time and space, are problematic since a greater proportion of cells with zero deaths in some age groups occur when we drill down to finer spatio-temporal units. Even with small numbers of deaths, variance estimates are unstable.
This is a small area estimation \citep{rao:molina:15} and  at the scale for which inference is desired, smoothing in space and time is required. 

In general, the design must be acknowledged when inference is performed, otherwise biased estimates with an incorrect measure of uncertainty will be produced. As an extreme example, in the DHS sampling is stratified by urban/rural and if a particular county only urban clusters were selected then ignoring this aspect will lead to bias in the estimation of the county level estimate, if U5MR is associated with urban/rural. As an alternative to design-based inference, a more traditional statistical approach may be employed in which a probability model for the observations is assumed, and the mean model contains terms that reflect the design, with a carefully chosen variance model. This approach is known as {\it model-based inference}; \cite{wakefield:simpson:godwin:16} compare the two approaches via simulation in a spatial context.



As in \cite{mercer:etal:15} we assume a discrete hazard model, with six hazards for each of the (monthly) age bands: [0,1), [1,12), [12,24), [24,36), [36,48), [48,60].
Detailed argument in \cite{allison:14} show that the contributions for a generic child correspond to the product of up to 60 Bernoulli likelihoods with $Y_{m,k}(\bms_j,t)$ being a binary indicator of survival in month $m$, $m=0,\dots,59$, for a child in survey $k$, in a household sampled at location $\bms_j$ in year $t$, $t=1980,\dots,2014$.
For a month beginning at $m$, the hazard  is,
$$\tensor*[_{1}]{q}{_{m,k}}(\bms_j,t)  = \frac{\exp[~\beta_{a[m],k}(\bms_j,t)~]}{1+\exp[~\beta_{a[m],k}(\bms_j,t)~]}=
\mbox{expit}[~\beta_{a[m],k}(\bms_j,t)~], 
$$
where $a[m]$ links the month $m$ to the six age bands $a$, i.e.,
$$
a[m] = \left\{
\begin{array}{ll}
1 & \mbox{if }m=0,\\
2 & \mbox{if }m=1,\dots,11,\\
3 & \mbox{if }m=12,\dots,23,\\
4 & \mbox{if }m=24,\dots,35,\\
5 & \mbox{if }m=36,\dots,47,\\
6 & \mbox{if }m=48,\dots,59.\\
\end{array}
\right.
$$
The likelihood for survival from month $m$ to $m+1$ in survey $k$ and at location $\bms_j$ in year $t$ is,
$$
Y_{m,k}(\bms_j,t) |  \tensor*[_{1}]{q}{_{m,k}}(\bms,t) \sim \mbox{Bernoulli}\left[~\tensor*[_{1}]{q}{_{m,k}}(\bms_j,t)~
\right].
$$
The latent logit model consists of a  part that is used for prediction, random effects to acknowledge the cluster sampling, survey  and independent temporal effects, and an offset that adjusts for the bias due to HIV epidemics, given in (\ref{eq:BIAS}). In summary, 
\begin{eqnarray}
\mbox{logit}[ \tensor*[_{1}]{q}{_{m,k}}(\bms_j,t) ] &=& \log (~ \mbox{BIAS}_{l[\scriptsize{\bms_j}],k }(t)~ ) + \beta_{a[m]}(\bms_j,t) + \eta_j + \upsilon_k+ \epsilon_t\nonumber \\
\beta_{a[m]}(\bms_j,t)&=&\beta_{a[m]}+\delta_{\text{str}[\scriptsize{\bms_j}]}
+   \phi_a(t)  + u(\bms_j,t) . \label{eq:pred}
\end{eqnarray}
We now describe each of the components. More details on the HIV bias offset are given in the supplementary materials, but the adjustment is carried out at the province level, indexed by $l$, with $l [{\bms_j}]$ corresponding to the province in which the cluster at $\bms_j$ is located. The random cluster effects $\eta_j \sim_{iid} \mbox{N}(0,\sigma^2_\eta )$ allowing for dependence amongst the individuals in the households at location $\bms_j$; hence, these effects will acknowledge the cluster design. The survey random effects $\upsilon_k \sim \mbox{N}(0,\sigma^2_\upsilon)$ allow for systematic biases in each of the three surveys (though of course this is relative to the average of the three surveys, and does not correct for any overall bias in the three surveys combined).  The temporal terms $\epsilon_t \sim_{iid}  \mbox{N}(0,\sigma^2_\epsilon)$ allow for yearly perturbations that have no structure in time. Each of the six age bands, has its own intercept $\beta_{a[m]}$.  The surveys are each stratified on an urban/rural indicator and on either 8 (years 2003 and 2008--2009) or 47 (year 2014) areas. The area-level stratification is strongly confounded with space and so we do not include a fixed effect for these strata, rather we assume the spatial field accounts for any such differences at a relatively large scale. The urban/rural classification changes far more quickly around urban centers, and for this reason we include a strata fixed effect $\delta_{\text{str}[\small{\bms_j}]}$. The temporal
 terms  $\phi_a(t)$  are random walks of order 2 (RW2), with one each for [0,1) and [1,12) months and then a third for the remaining period of [12,60] months. We decided on these splits based on initial analyses and on the demographic pattern in which the majority of U5MR deaths occur in the first year of life. In each, for reasons of parsimony, the same precisions were used (we investigated the use of different precision parameters for the three age groups, but there was little difference in the resultant inference), i.e.,~the distribution is $\mbox{RW2}(\sigma_\phi^2)$ for all three age bands. Sharing the precision parameter forces the same smoothness in the temporal evolution for the logit of the hazard in 
each age group, but the temporal trends are independent between age groups,
conditional on the precision parameter. The RW2s have sum-to-zero constraints to make them identifiable when combined with the age-group-specific intercepts.The most complex term to explain is the space-time interaction $u(\bms,t)$, and we begin with a description of separable processes.

A separable spatio-temporal process has a covariance function that is a combination of a spatial 
dependence structure, $c_\mathrm{S}$, and a temporal dependence structure, $c_\mathrm{T}$, through 
$$
	c_\mathrm{ST}(~(\boldsymbol{s}_1, t_1), (\boldsymbol{s}_2, t_2)~) = c_\mathrm{S}(\boldsymbol{s}_1, \boldsymbol{s}_2)\times c_\mathrm{T}(t_1, t_2), \qquad \text{for all $t_1$, $t_2$, $\boldsymbol{s}_1$ and $\boldsymbol{s}_2$}.
$$
The multiplicative structure is beneficial because it is easy to construct valid
spatio-temporal covariance functions by combining valid spatial and temporal
covariance functions.

In our model we use a combination of a Mat\'ern covariance structure for space, which is approximated via SPDE,
and an AR(1) process in time. Inference is done using INLA with samples drawn from the approximate posterior for inference on functions of interest.  The process is written as $u(\boldsymbol{s},t)$
and is a combination of a temporal structure $c_\mathrm{T}$ and a spatial structure, $c_\mathrm{S}$
which translates to,
$$
\Sigma_\text{ST} = \Sigma_\text{T} \otimes \Sigma_\text{S},
$$
if the process is observed on $(\boldsymbol{s}, t)\in\{\boldsymbol{s}_1, \ldots, \boldsymbol{s}_N\}\times \{1, 2, \ldots, T\}$ (in which case $\Sigma_\text{S}$ is $N \times N$,  $\Sigma_\text{T}$ is $T \times T$ and  $\Sigma_\text{ST}$ is $NT \times NT$).

The hazard for each age group is expected to vary spatially, but due to data sparsity
the data will not support a separate spatial main effects for each of the six age bands. A parsimonious
model would include a shared spatial main effect for the age groups, but since a 
spatio-temporal interaction is necessary to account for the yearly changes in the spatial pattern, we do not include the spatial main effect. It is too expensive to apply the necessary 
temporal sum-to-zero constraints that would be required to give identifiable spatial main effects
alongside a spatio-temporal interaction.Therefore, the shared spatial main effect and the shared spatio-temporal interaction are both handled with a separable spatio-temporal model that combines an AR(1) structure with the Mat\'ern covariance function. 
The resulting spatio-temporal covariance function
can be explained through a constructive example which gives some intuition on the space-time interaction. A stable AR(1) process with
marginal variance 1 can be generated by
$$
	a_{t+1} = \rho a_t + \epsilon_t, \qquad t = 2, 3,\ldots, T,
$$
where $\epsilon_i \sim_{iid} \mbox{N}(0, (1-\rho^2))$, for $i = 2, \ldots, T$, and
$a_1\sim \mbox{N}(0, 1)$. The temporal process can be made spatio-temporal by
replacing the starting condition and the innovations with spatial Mat\'ern fields, to give
$$
	a_{t+1}(\boldsymbol{s}) = \rho a_t(\boldsymbol{s}) + \epsilon_t(\boldsymbol{s}), \qquad i = 2, 3, \ldots, T,
$$
where $\epsilon_i \sim \mbox{N}(0, (1-\rho^2)c_\mathrm{S}(\cdot))$, and
$a_1 \sim \mbox{N}(0, c_\mathrm{S}(\cdot))$, where $c_\mathrm{S}$ is the
stationary Mat\'ern covariance function.  Hence,  a proportion $\rho^2$ of the marginal variance is explained by the previous time step and a proportion $1-\rho^2$ is arising from a new realization of a spatial field.

The joint
identifiability of the six temporal trends and the spatio-temporal interaction can be achieved
through integrate-to-zero constraints for each year. This integration is carried out with respect to the spatially varying population density
$d(\bms)$:
$$
	\int u(\boldsymbol{s}, t) d(\bms) ~ \mathrm{d}\boldsymbol{s} = 0, \qquad t = 1980, \ldots, 2014,
$$
where $u(\boldsymbol{s}, t)$ is the separable spatio-temporal process. These yearly integrate-to-zero
constraints mean that the spatial average of the spatio-temporal effect is constantly equal to 
zero and that the temporal change in the spatial average of the logits of the hazards of each age
group is explained by the corresponding temporal main effects. In particular, the RW2 trends are approximately interpretable as the change in the national level with time.

This spatio-temporal effect on a temporal resolution of 35 years is too expensive
to include in the Bayesian model, but since
we want the spatio-temporal process to change gradually in time, it is possible
to use an approximation that changes piecewise linearly in time, a similar approach was taken in \nocite{blangiardo:cameletti:15}Blangiardo and Cameletti (2013, Chapter 8). We decrease the 
resolution of the spatio-temporal process to 8 time steps by defining $\widetilde{u}_h(\boldsymbol{s})$
for knot locations $h= 1, 2, \ldots, 8$, corresponding to years $1980, 1985, \ldots, 2015$, and defining
$$
	u(\boldsymbol{s}, t) = (1-\alpha_k(t))\widetilde{u}_h(\boldsymbol{s}) + \alpha_h(t)\widetilde{u}_{h+1}(\boldsymbol{s}), \qquad \text{for $1975+5h \leq t < 1980+5h$},
$$
where $\alpha_h(t) = t/5 - \mbox{floor}(t/5)$ gives the factor required for linear interpolation between
the two knot locations. Note that if the integrate-to-zero constraint is satisfied for 
$\widetilde{u}_h(\boldsymbol{s})$ for $h= 1, 2, \ldots, 8$, the integrate-to-zero constraint is
also satisfied for linear combinations $u(\boldsymbol{s}, t)$ for $t=1980,1981,\dots,2015$.

Each of the precisions for the independent and identically distributed effects have Gamma(0.5, 0.0005) priors (which give 5\%, 50\%, 95\% quantiles for the standard deviations of 0.016, 0.047, 0.52). The spatial part of the spatio-temporal interaction has a ``penalized complexity"  prior \citep{fuglstad:etal:15,simpson:etal:17} with $\Pr(\text{ spatial range } < 0.5) = 5\%$ and $\Pr(\text{ spatial $\sigma$} > 3) = 5\%$; all other parameters have default priors.

For predictions, the 
cluster, survey and temporal independent and identically distributed effects are not included so that the only contribution is $\beta_a(\bms_j,t)$.
The predicted U5MR at location $\bms_j$ and at time $t$ is, 
\begin{eqnarray*}
\mbox{U5MR}(\bms,t) &=& 
1- \prod_{a=1}^6
\left[ \frac{1}{1+\exp[\beta_a(\bms,t)]}\right]^{z[a]},
\end{eqnarray*}
where $z[a]=1,11,12,12,12,12$, for $a=1,\dots,6$ and with $\beta_a(\bms,t)$ given by (\ref{eq:pred}).

The data and the fitted model are on a continuous spatial scale, but the aim is to produce
values on a discrete scale using the 47 administrative regions. The information available is
the posterior of the spatially varying U5MR and the population density
$d(\bms)$. We obtained the latter from \url{worldpop.org} \citep{linard:etal:12}. We really should be using the births density, but such data are difficult to obtain; we examined a surface of estimated live births for one year that was available \citep{worldpop:births:17}, and the surface for that year showed very little change. We assume that the infinite superpopulation has the same relative variation in population
density as the real population and define the U5MR of region $i$ by
\begin{equation}\label{eq:agg}
	\text{U5MR}_i(t)= \frac{\int_{R_i} \text{U5MR}(\bms,t)  d(\boldsymbol{s}) \, \mathrm{d}\boldsymbol{s}}{\int_{R_i} d(\boldsymbol{s}) \, \mathrm{d}\boldsymbol{s}}, \quad i = 1, 2, \ldots, 47,
\end{equation}
where $R_i$ denotes administrative region $i$. This averaging gives zero weight to areas with no population, even though the continuous surface is defined at such points.


\subsection{Constructing a Space-Time Surface Results}

\begin{figure}
\centering
\includegraphics[width=0.4\linewidth]{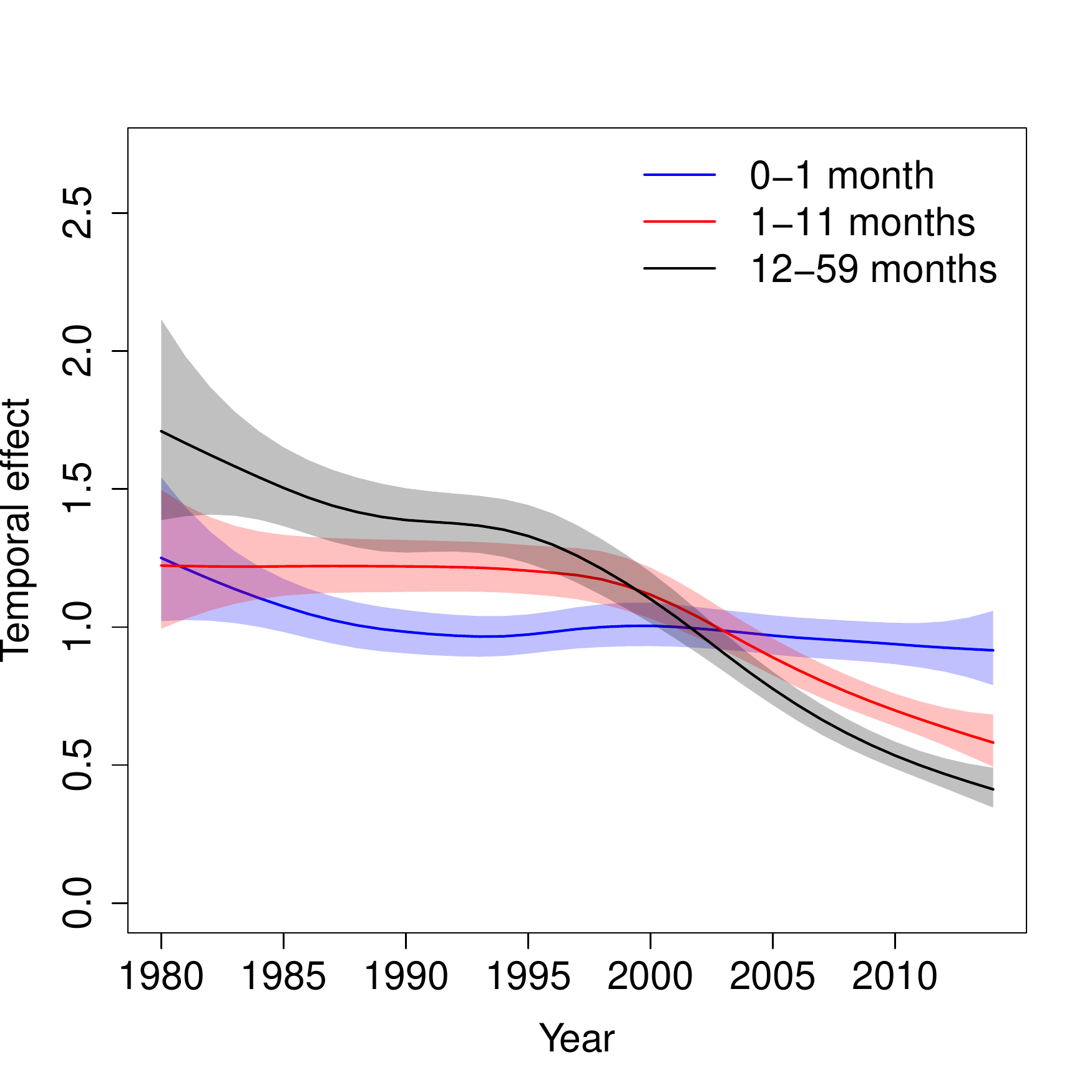}
\includegraphics[width=0.4\linewidth]{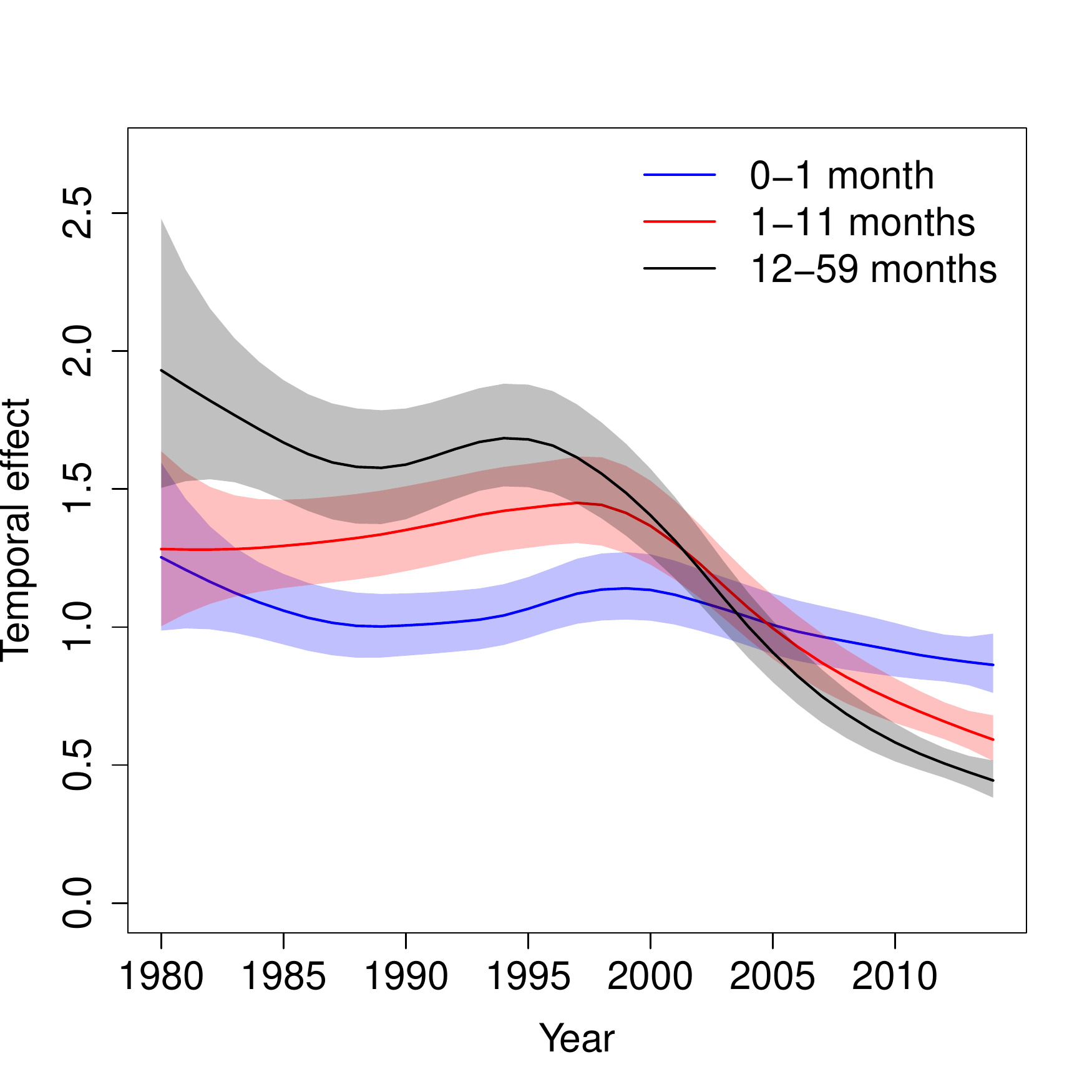}
\caption{Left: Median RW2 model temporal trends (left) HIV adjusted time trends (right) for the  three age bands.  Both with 95\% pointwise credible intervals.}
\label{fig:RW2}
\end{figure}

We begin by summarizing inference on some of the key elements of the model, before reporting on
substantive summaries. The left panel of Figure \ref{fig:RW2}  shows the posterior medians of the RW2 median fits for each of the
[0,1), [1,12), [12,60] age groups, along with 95\% point-wise credible interval envelopes. We see that the temporal trend decreases for all three age groups. While the [0-1] age group shows a decreasing slope from 1995 onwards, a continuing strong decrease can be seen for the other two age groups, with the most prominent drop being for the 12--59 month age group. One of the major reasons for the drop in U5MR is increased vaccination \citep{haakenstad:etal:16}, but this is carried out once maternal antibodies are no longer present after 6--12 months. The right panel of Figure \ref{fig:RW2}  shows the HIV adjusted version of this plot and the effect of the epidemic is clear to see in all three age groups.

 Table \ref{tab:postresults} gives posterior summaries of key parameters in the space-time model. The standard deviations are not all comparable since for the RW2 it is a conditional standard deviation. The spatio-temporal standard deviation is relatively large indicating that there are strong spatial effects for the Kenya data and the median of the range parameter is 1.77$\degree$, which is quite large (about a fifth the size of the study region). There is also strong year-to-year correlation in the AR(1) model.

\begin{table}[htp]
\begin{center}
\begin{tabular}{l|l|l|l}
Parameter& 2.5\%&50\%&97.5\%\\ \hline
Standard deviation for RW2 time & 0.0097 & 0.018 & 0.031\\
Standard deviation for IID-time         &                  0.023  &  0.049  &    0.099\\
Range for spatio-temporal effect         &                 1.28&  1.77&   2.45\\
Standard deviation for spatio-temporal effect &            0.49&  0.59& 0.71\\
AR(1) parameter for spatio-temporal effect &     0.78&   0.86&    0.92\\
Standard deviation for IID-cluster          &              0.32& 0.36&     0.39\\
Standard deviation for IID-survey            &             0.017   & 0.045  &    0.13\\
Effect of rural vs urban                               &   0.011   & 0.080  &    0.15\\ 
\end{tabular}
\end{center}
\caption{Posterior quantiles for model parameters.}
\label{tab:postresults}
\end{table}%

\begin{figure}
\centering
\includegraphics[width=0.5\linewidth]{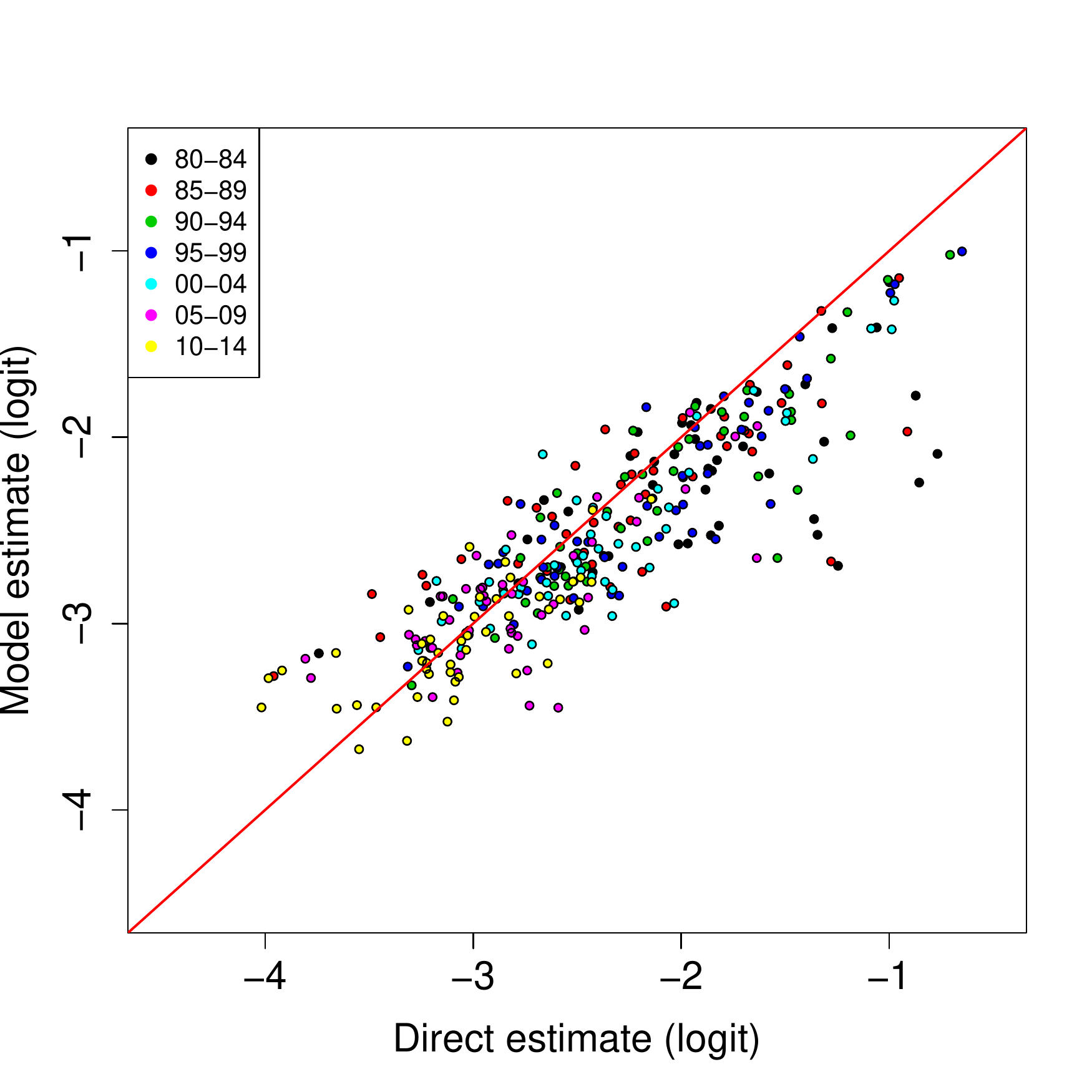}
\caption{Modeled estimates versus weighted (direct) estimates on the logit scale, color coded by period..}
\label{fig:directcomp}
\end{figure}

Figure \ref{fig:directcomp} shows a comparison between the modeled $\qf$ and weighted estimates at the 47 county level, and aggregated over 5 years (aggregation over years is required, otherwise the direct estimates are unstable). We see some attenuation due to shrinkage, as expected. In the Supplementary Materials we include more detailed plots and show the uncertainty in the modeled and weighted estimators. Again, as expected, the modeled estimates have much greater precision.

\begin{figure}
\centering
\includegraphics[width=0.3\linewidth]{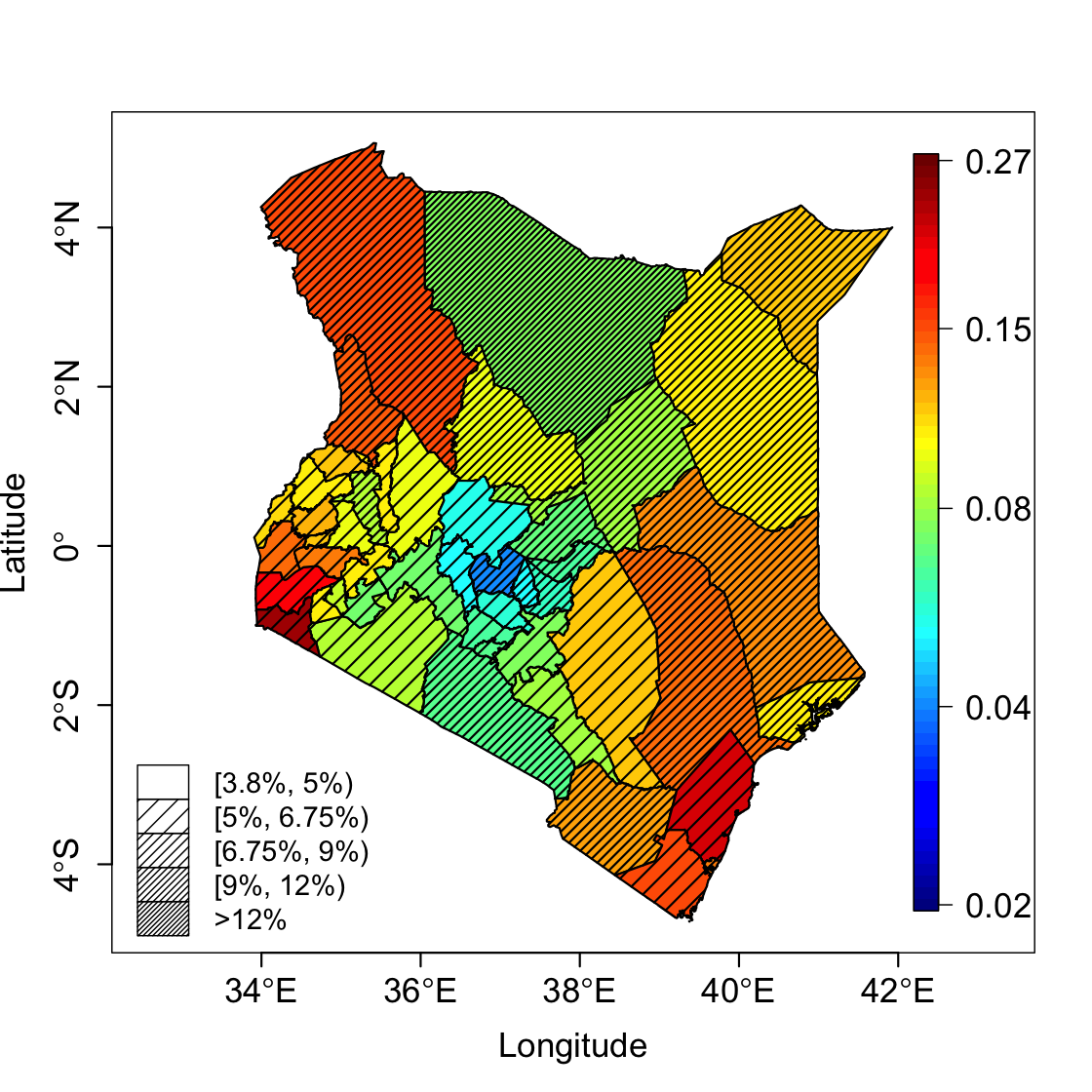}
\includegraphics[width=0.3\linewidth]{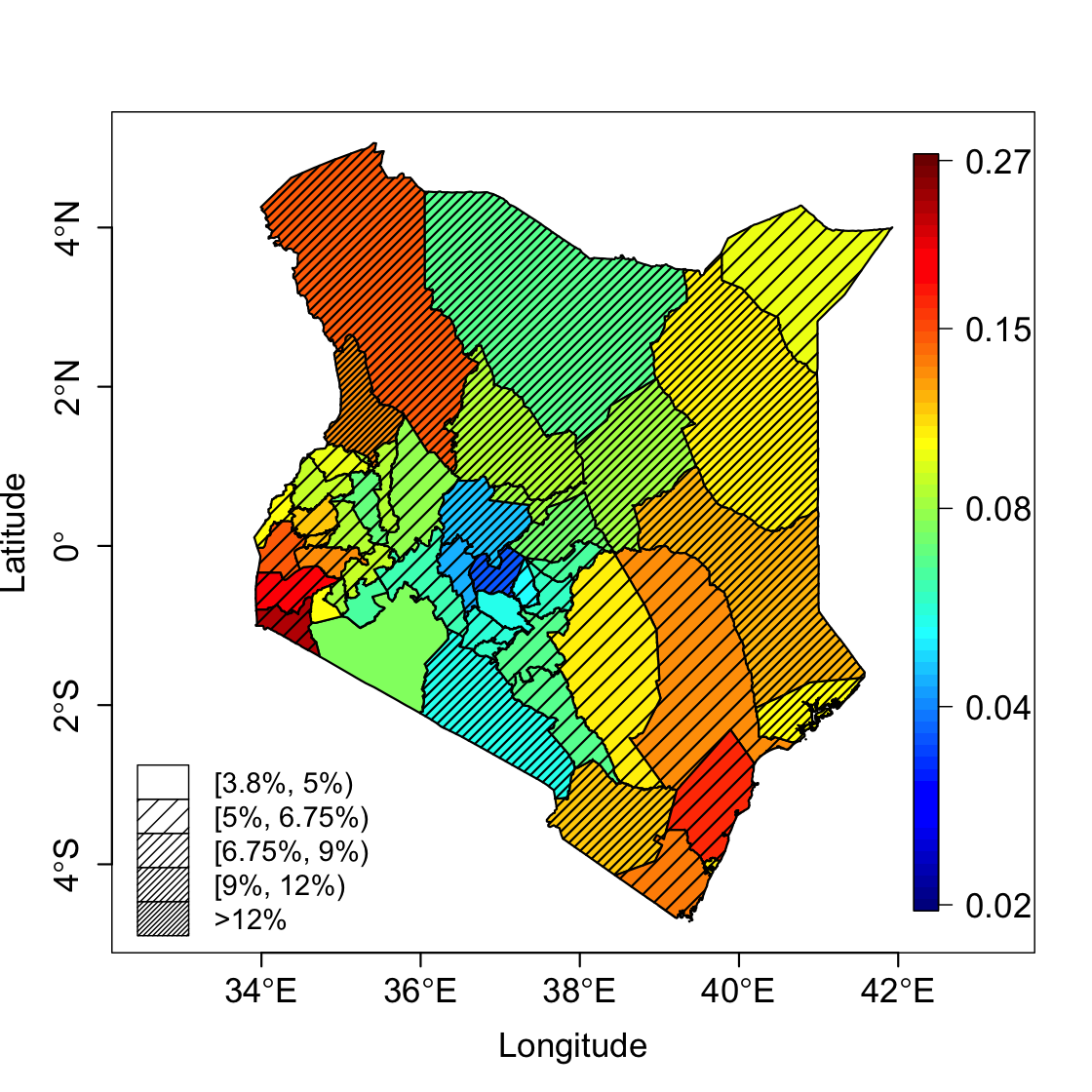}
\includegraphics[width=0.3\linewidth]{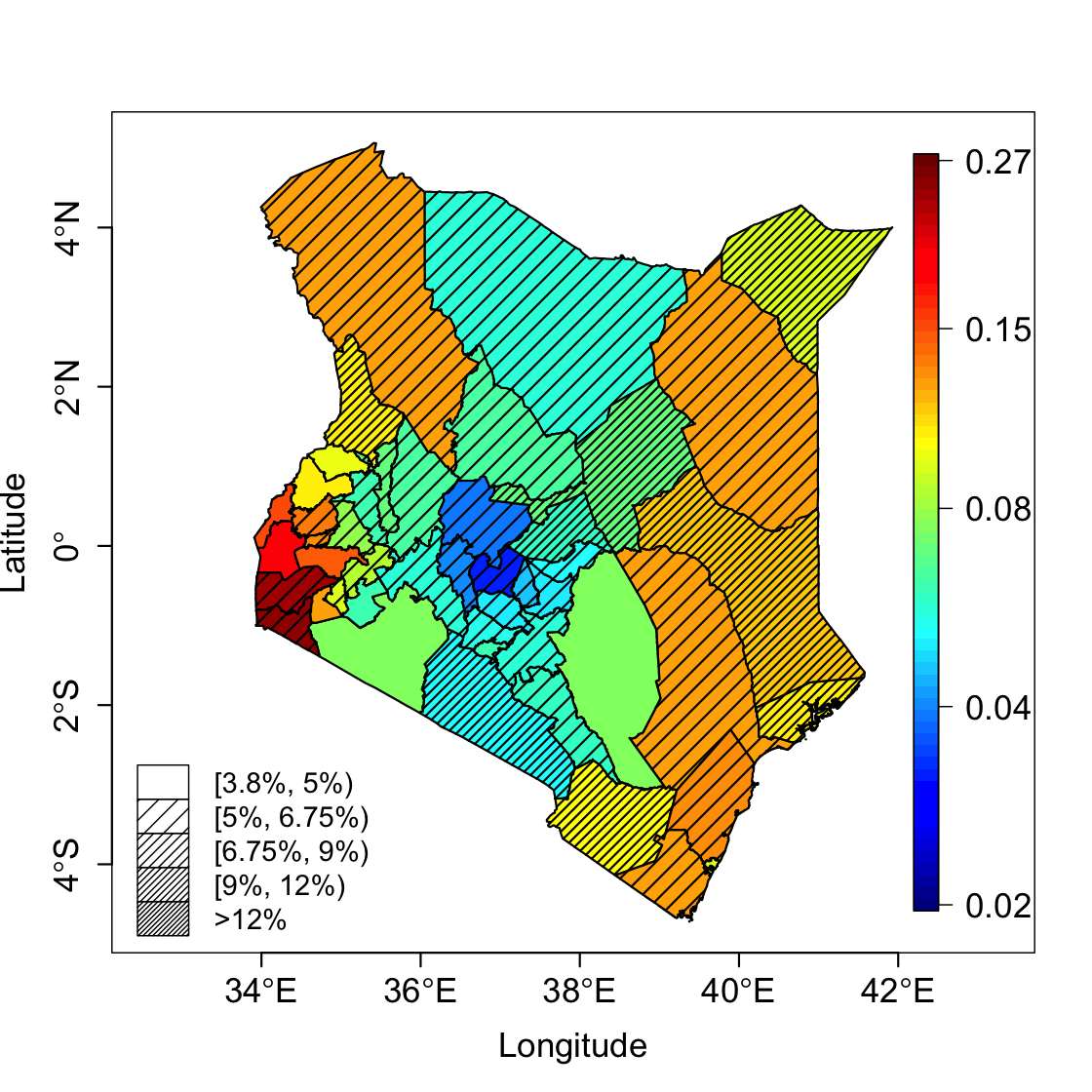}
\includegraphics[width=0.3\linewidth]{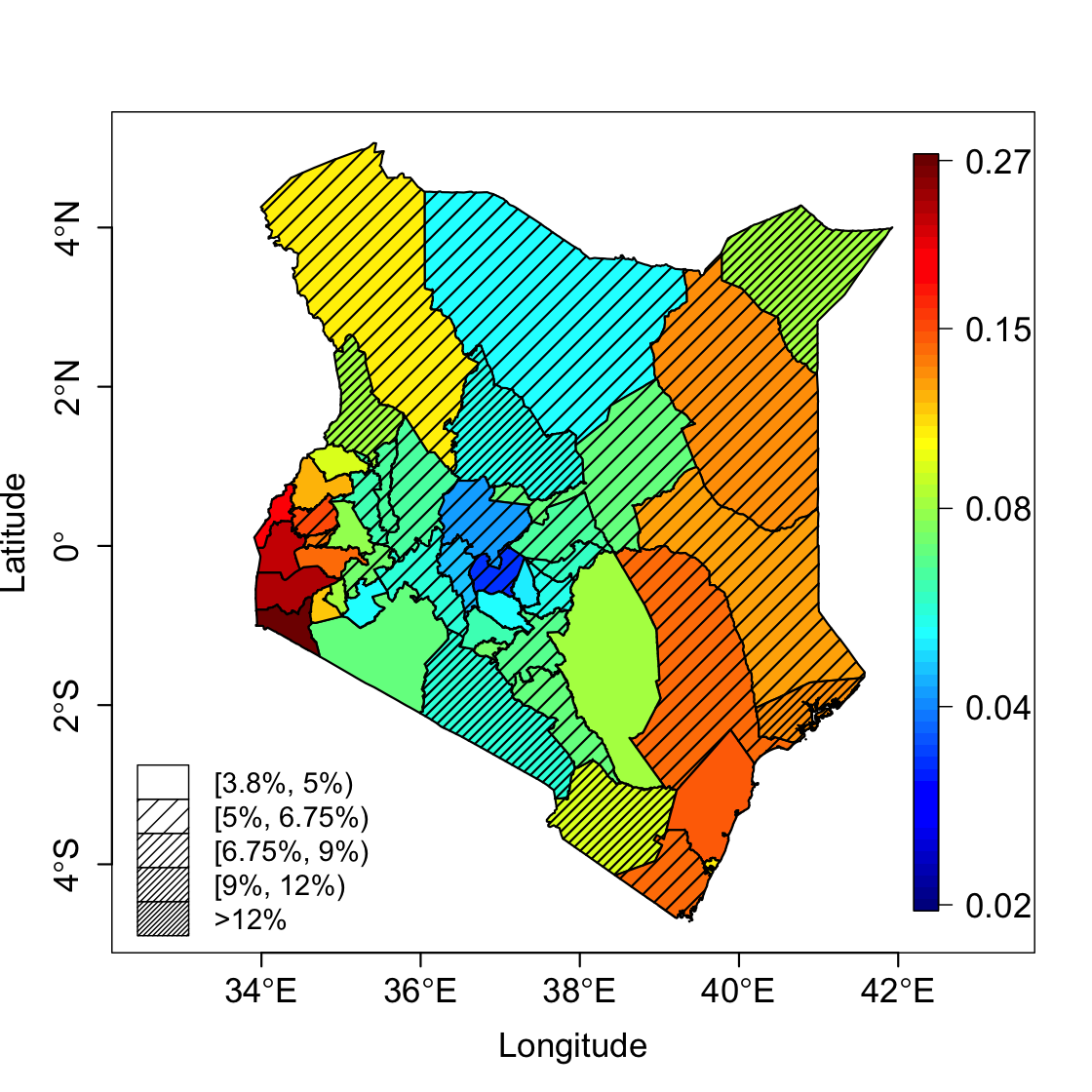}
\includegraphics[width=0.3\linewidth]{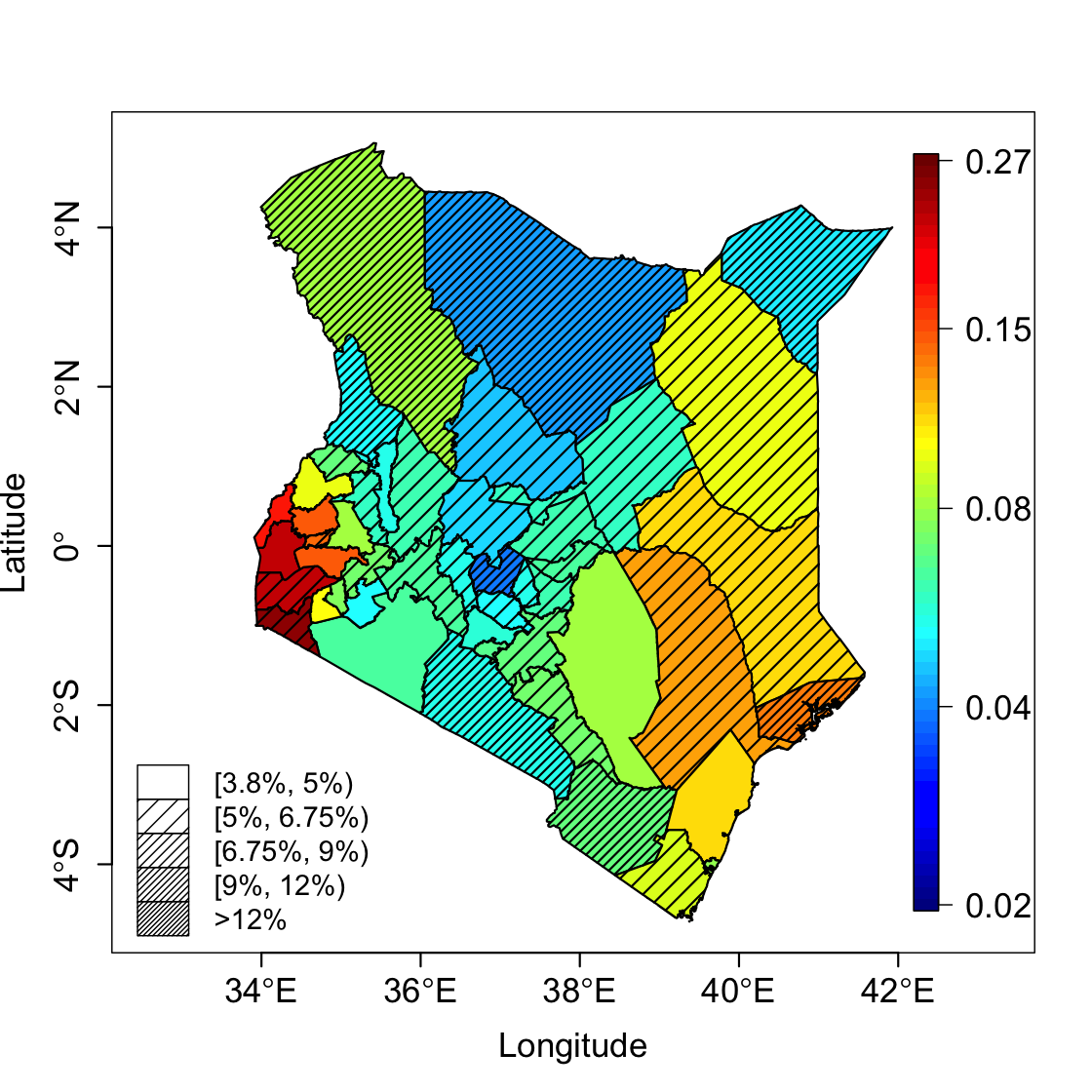}
\includegraphics[width=0.3\linewidth]{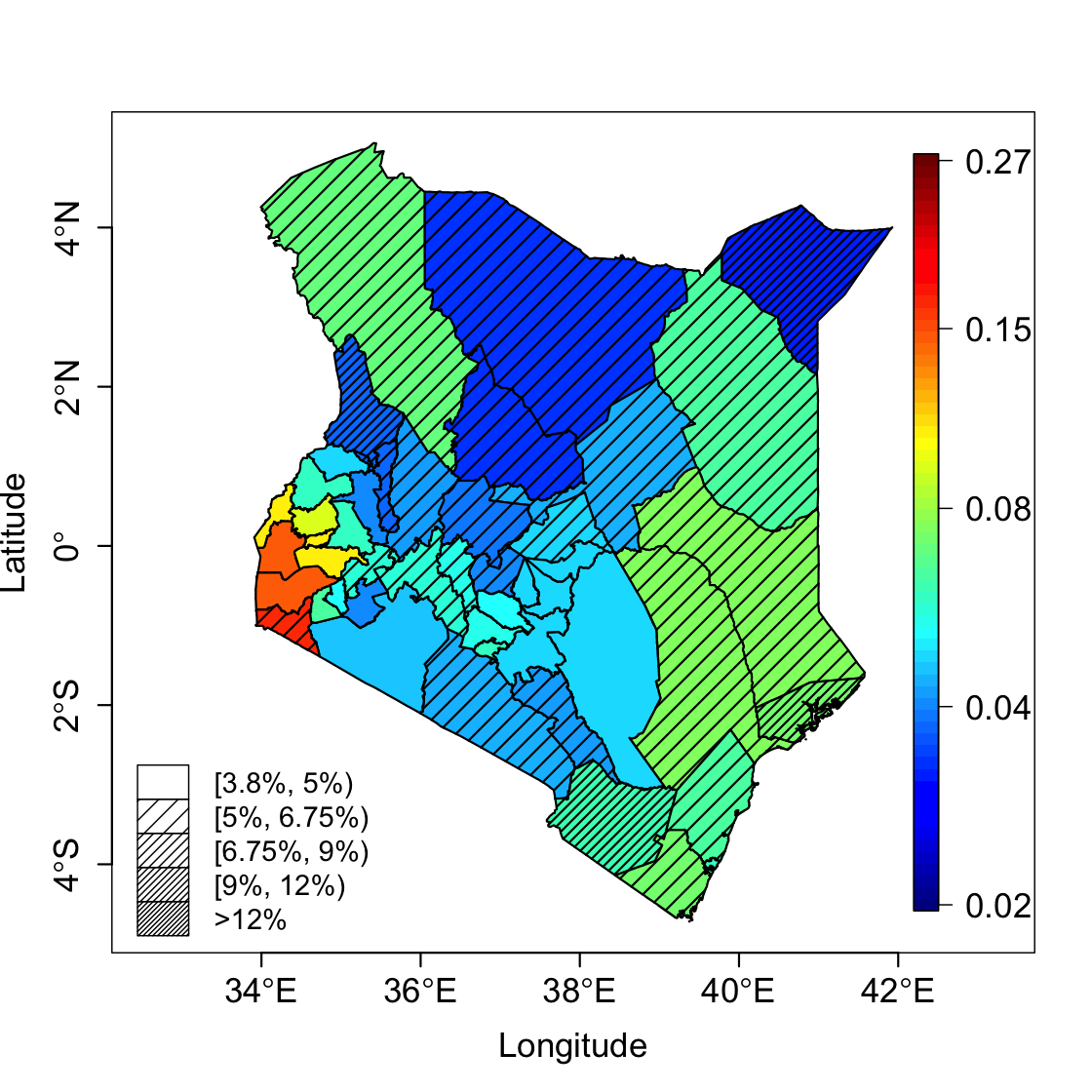}
\includegraphics[width=0.3\linewidth]{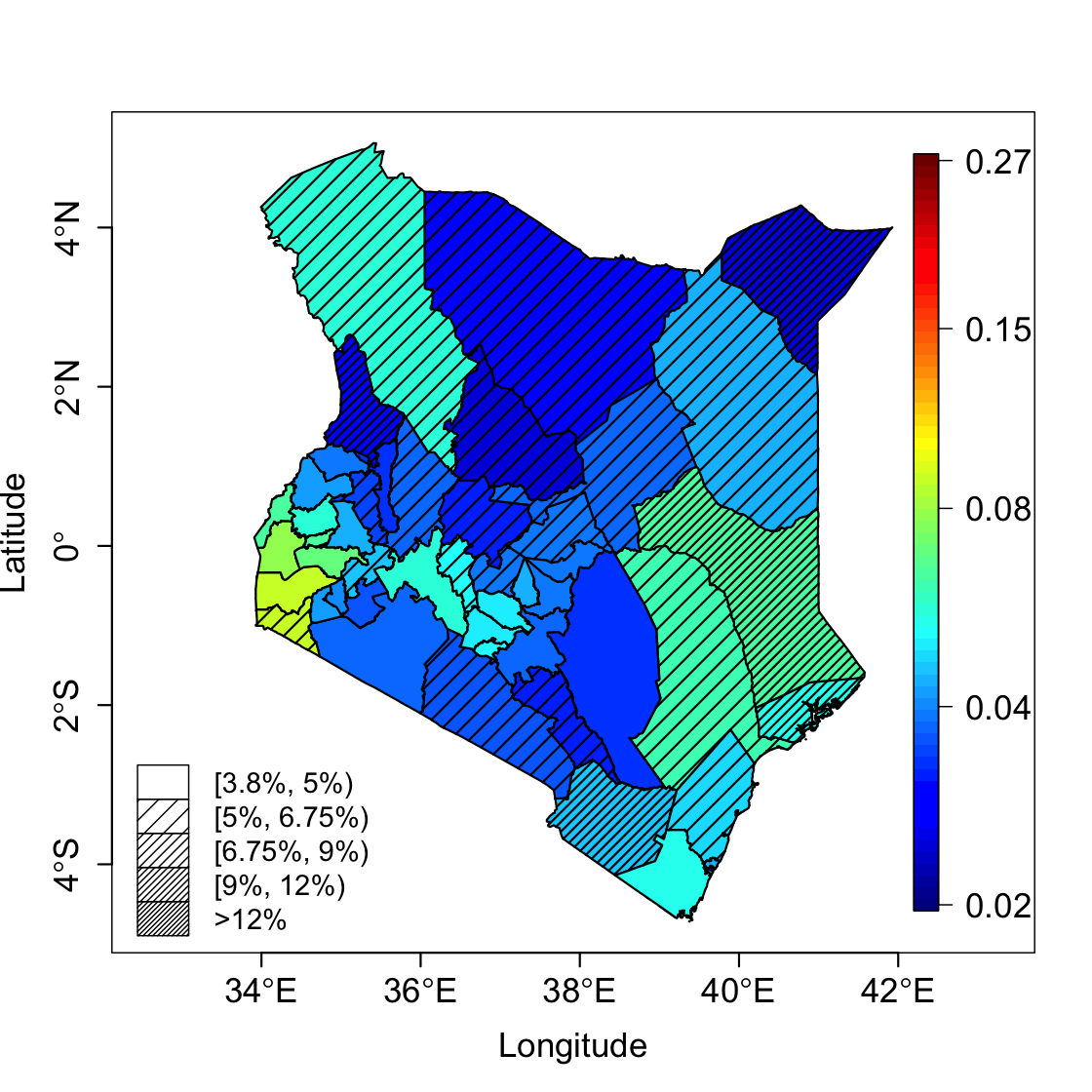}
\includegraphics[width=0.3\linewidth]{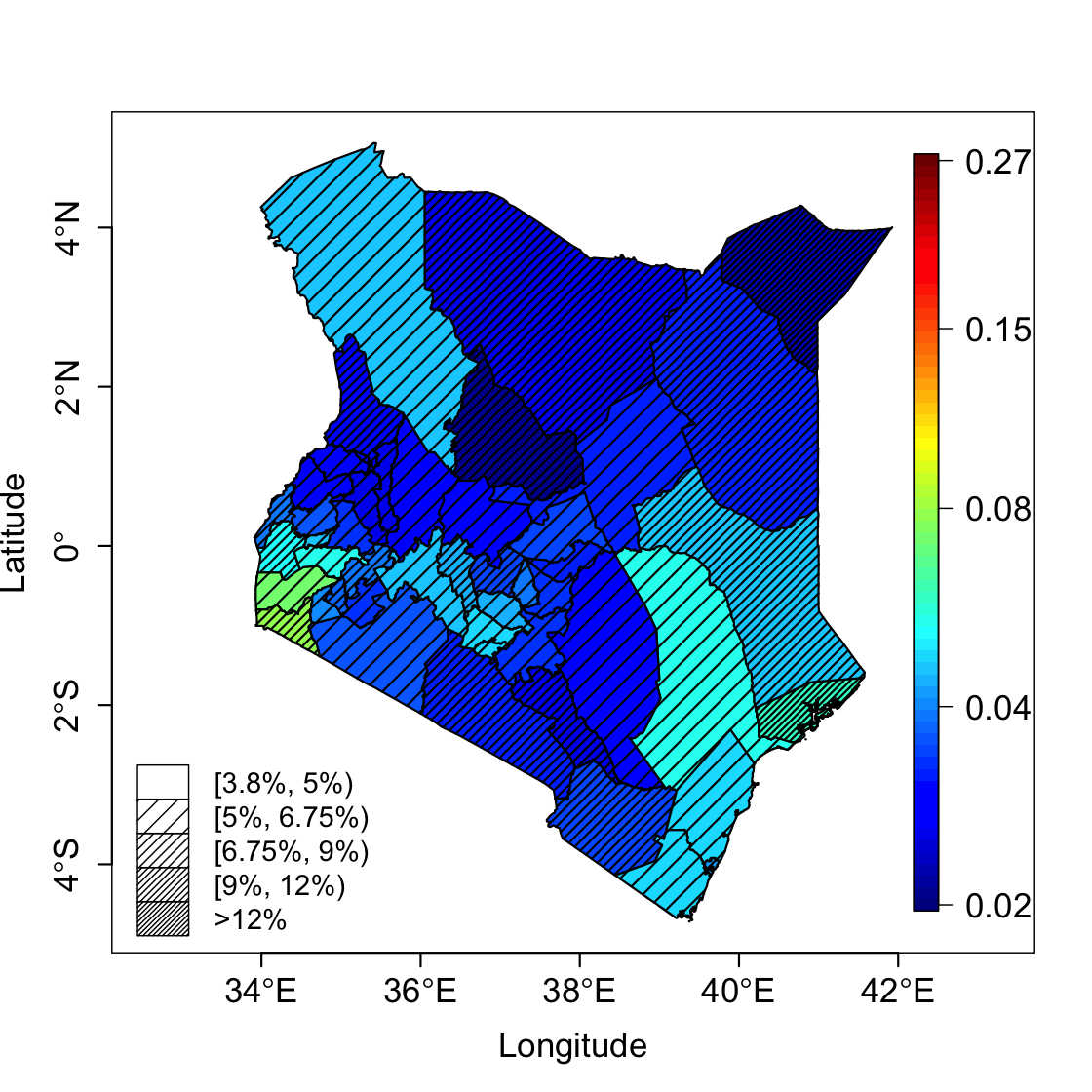}
\includegraphics[width=0.3\linewidth]{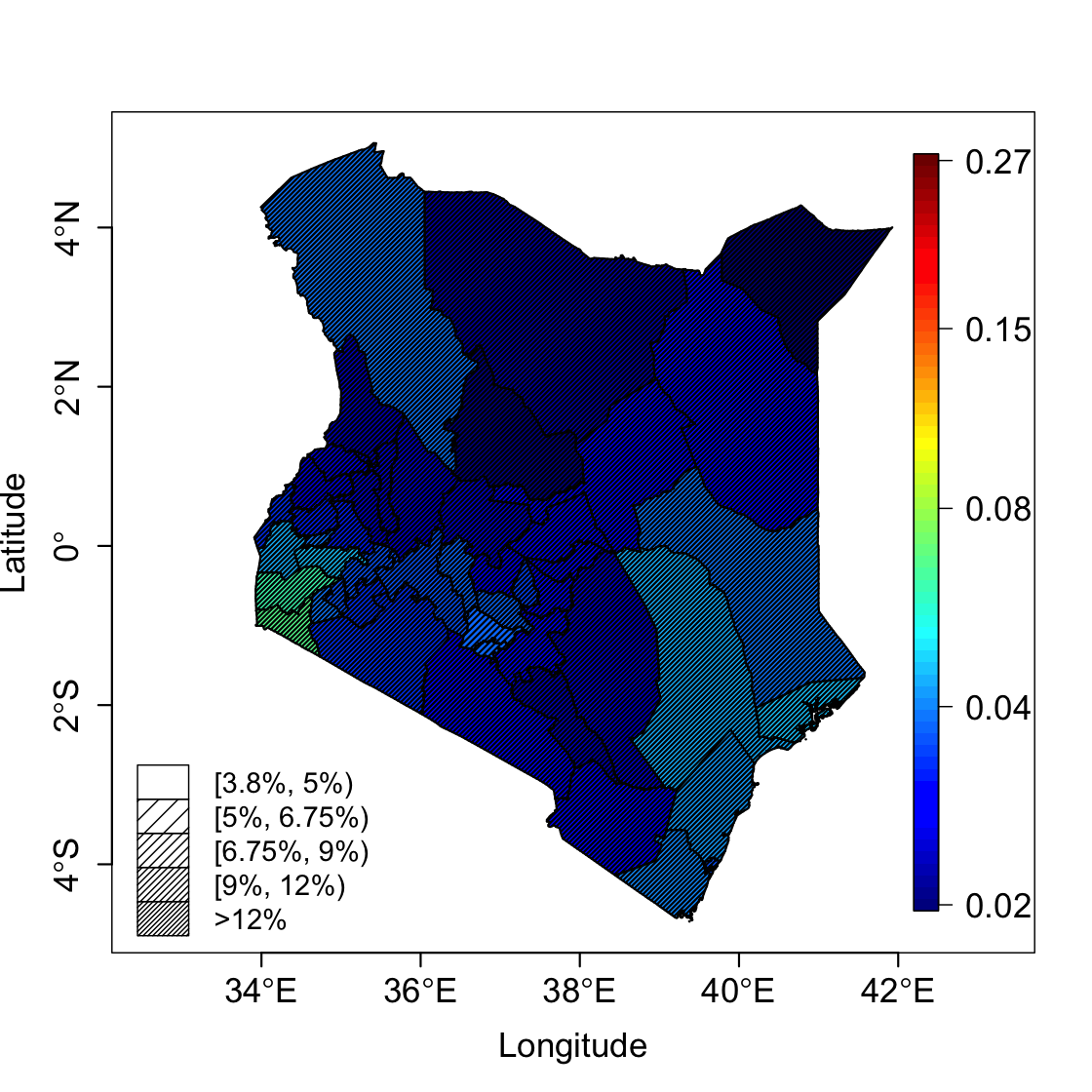}
\caption{Maps of the posterior median estimates of U5MR at the county level, with uncertainty represented by hatching. Top row: 1980, 1985, 1990. Middle row: 1995, 2000, 2005. Bottom row: 2010, 2015, 2020.}
\label{fig:5yearmaps}
\end{figure}

As mentioned in Section \ref{sec:introduction}, we wish to make inference at the spatial level at which policy interventions occur. For Kenya, this is at the 47 county level, and Figure \ref{fig:5yearmaps} shows a sequence of 9 maps of $\qf$ for the years $1980, 1985,\dots, 2015, 2020$ (we have 35 yearly estimates, but for space reasons we take a 5-year spacing). The last two of these years are obtained by forecasting from the model. The density of hatching reflects the uncertainty. The dramatic decrease over time in $\qf$ is apparent, though strong subnational variation persists. The Supplementary Materials contain maps of the uncertainty.

\begin{figure}
\centering
\includegraphics[width=0.3\linewidth]{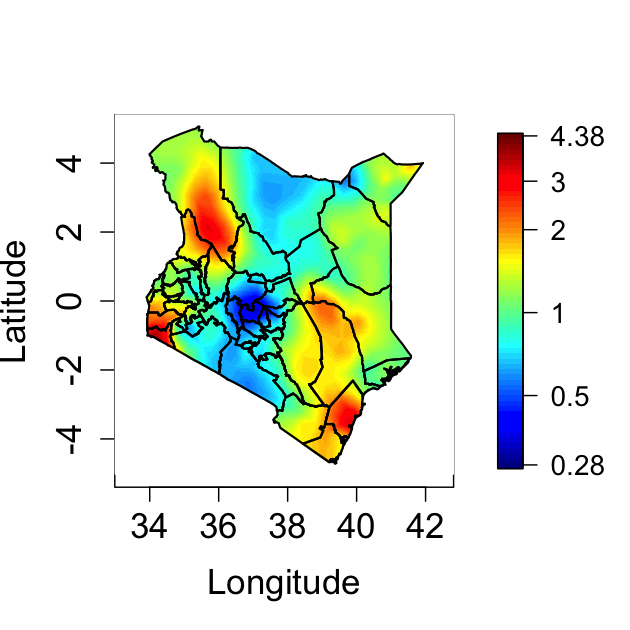}
\includegraphics[width=0.3\linewidth]{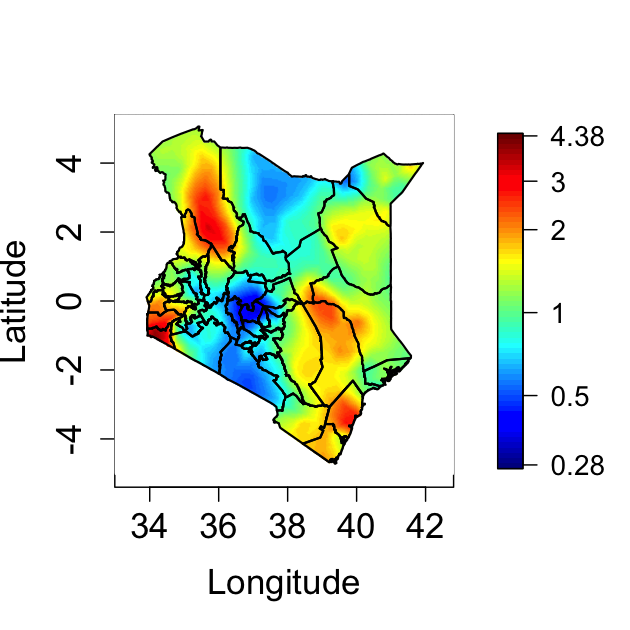}
\includegraphics[width=0.3\linewidth]{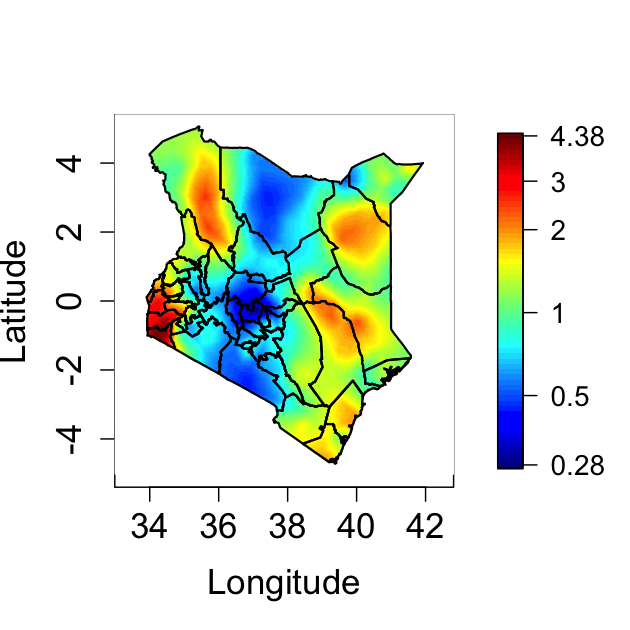}
\includegraphics[width=0.3\linewidth]{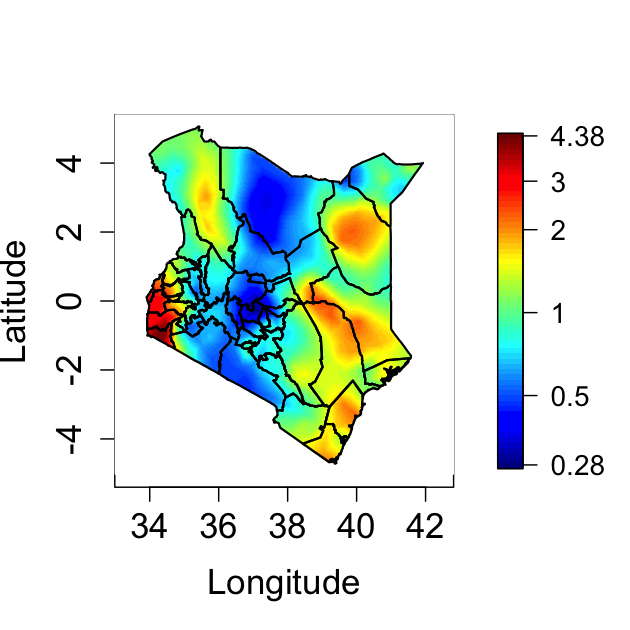}
\includegraphics[width=0.3\linewidth]{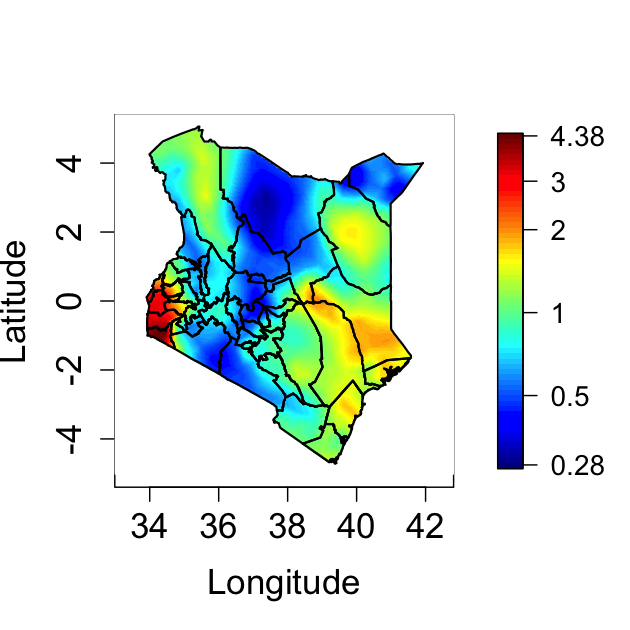}
\includegraphics[width=0.3\linewidth]{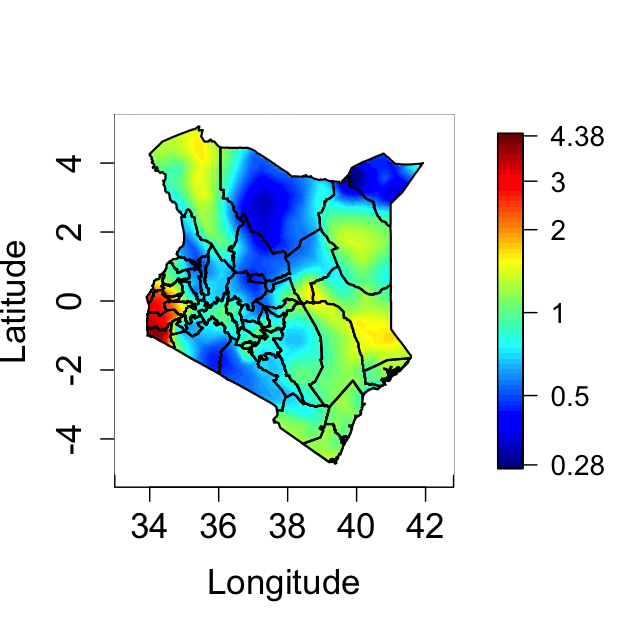}
\includegraphics[width=0.3\linewidth]{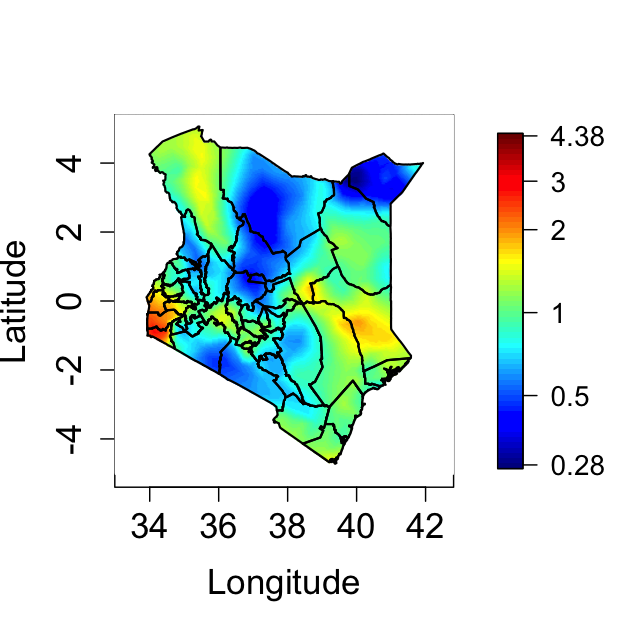}
\includegraphics[width=0.3\linewidth]{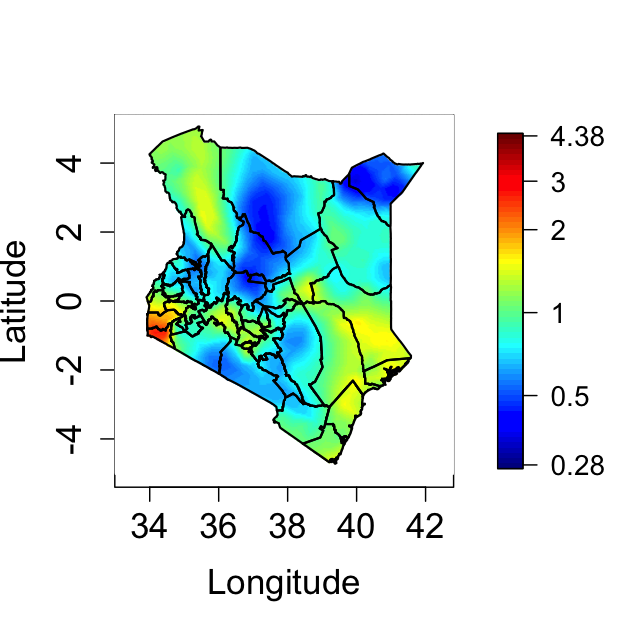}
\includegraphics[width=0.3\linewidth]{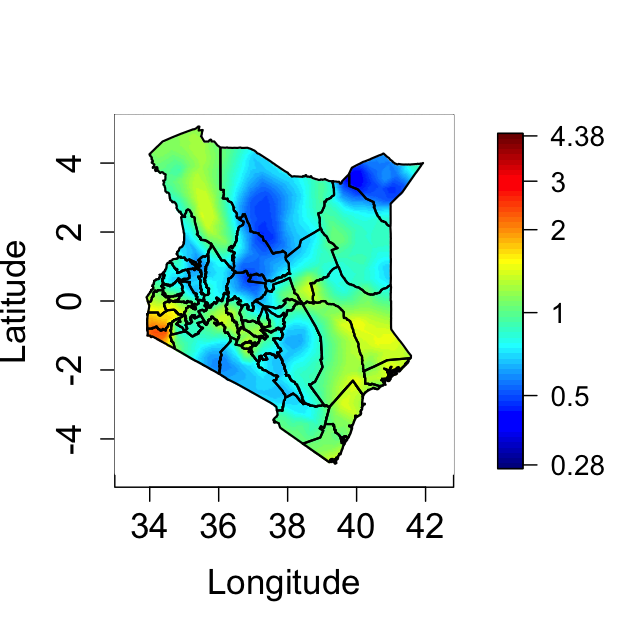}
\caption{Maps of the spatio-temporal odds surface, $\exp[~u(\bms,t)~]$. Top row: 1980, 1985, 1990. Middle row: 1995, 2000, 2005. Bottom row: 2010, 2015, 2020.}
\label{fig:spatioTempEffect}
\end{figure}

Figure \ref{fig:spatioTempEffect} shows the posterior medians of the spatio-temporal terms $\exp[ u(\bms),t) ]$ for the years 1980, 1985, 1990, \ldots, 2015, 2020. The last two of these years are obtained by predicting forward the space-time field. From 1980 onwards strong spatial effects can be seen in the counties Turkana and West Pokot in the north west part, the province Nyanza in the middle west part and the counties Kilifi, Tana River and Garissa in the south east part of Kenya. While the highs in the north west and south east have almost disappeared by 2004, a higher effect in the counties Minori and Homa Bay of the province Nyanza persist and, without interventions, one would expect these trends to continue until 2020. Around 1990 to 1995 higher effects can also seen in the north east.  

While it seems that the spatio-temporal trend decreases over time it should be emphasized that there is still a strong effect present in recent periods and also in the future. To illustrate this we computed the 95\% and 5\% points of the pixel values for each of the nine maps. In 1980 the $95\%$ quantile was $2.2$ and the $5\%$ quantile was $0.63$ leading to a ratio of $3.4$. While the $5\%$ quantile decreases until 2005 and then increases again, the $95\%$ quantile decreases almost constantly. The ratio of 95\% to 5\% points increases until 1995 with a value of $4.4$  and then decreases. In 2010 the ratio is still $3.5$ and in the predicted years $2015$ and $2020$ it is $2.97$ and $2.55$, respectively. Thus, there remains strong subnational differences in U5MR.

\begin{figure}
\centering
\includegraphics[width=0.4\linewidth]{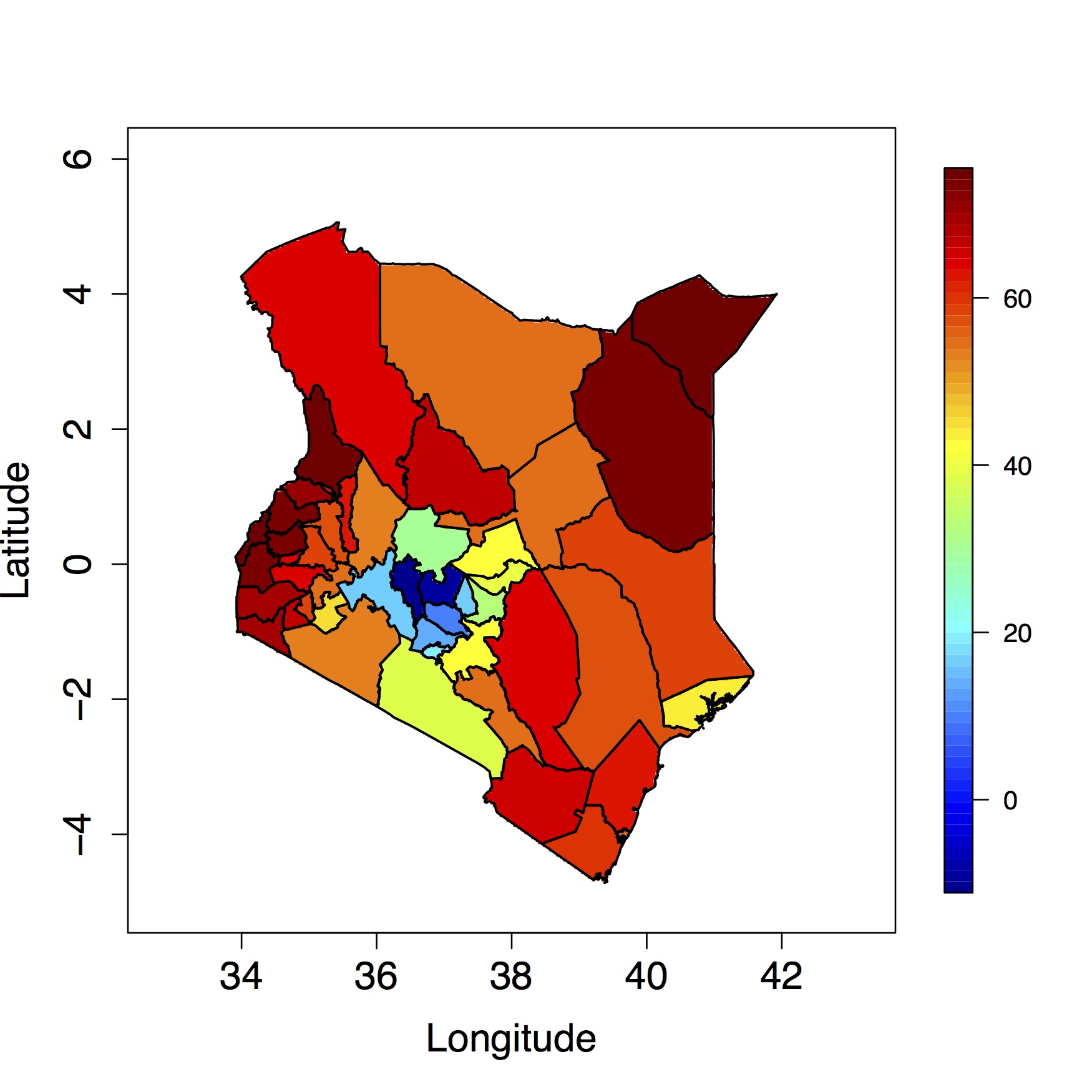}
\includegraphics[width=0.4\linewidth]{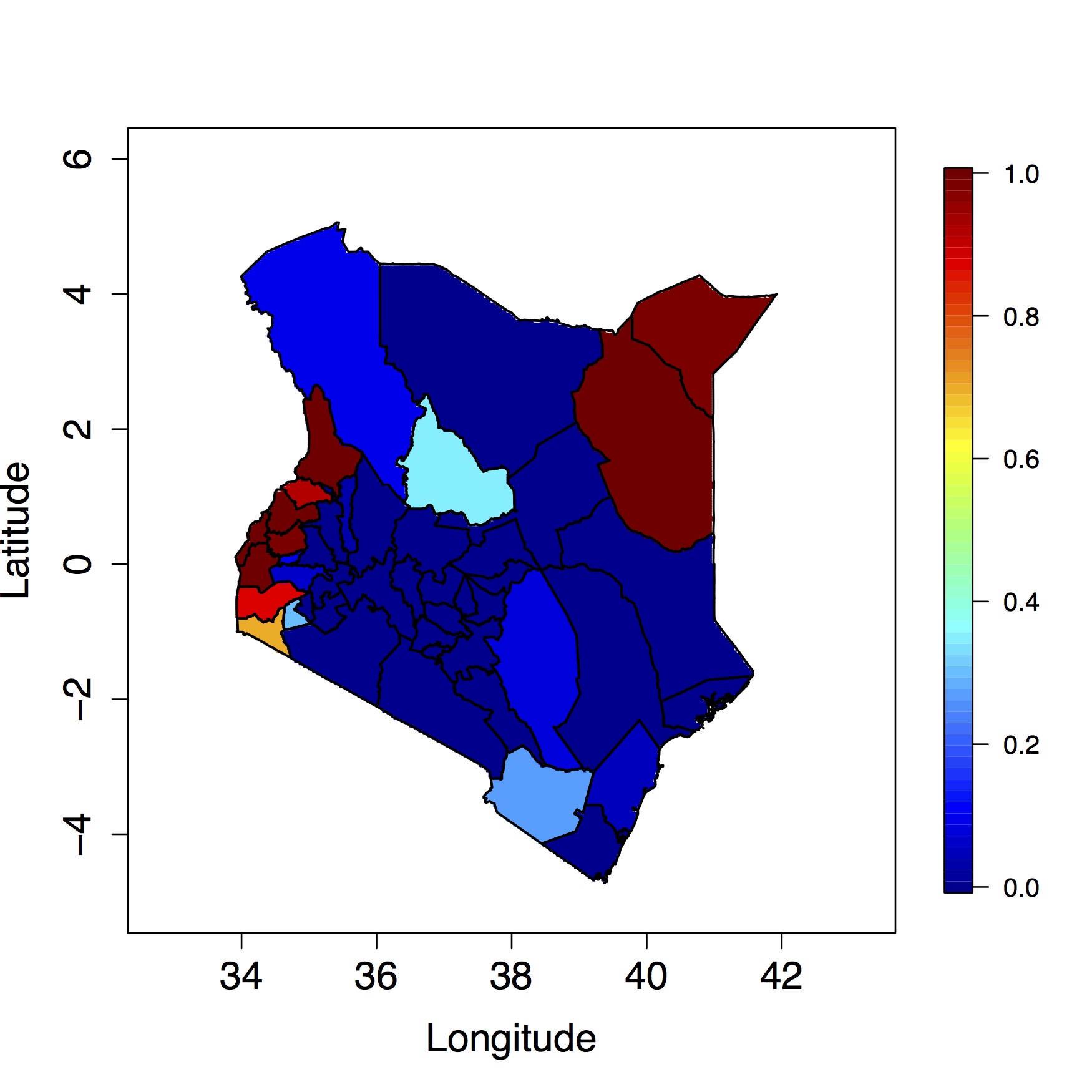}
\caption{Left plot: Posterior median of $100 \times [~\mbox{U5MR}_i(1990)-\mbox{U5MR}_i(2015)~]/\mbox{U5MR}_i(1990)$, right plot: Posterior probability of county $i$ achieving a 67\% drop over 1990--2015, $i=1,\dots,47$.}
\label{fig:MDG}
\end{figure}

The Millenium Development Goals  (MDG) aimed for a drop of 67\% in U5MR between 1990 and 2015. In the left hand panel of Figure \ref{fig:MDG} we map the posterior median of the percentage drop at the county level, with counties in the central part of Kenya experiencing very small decreases only. In the right hand panel we  plot the posterior probability that each county achieved this aim and we see  that very few attained a 67\% drop. Over the country the posterior median drop was 55\% with 95\% credible interval of (49.9\%, 60.0\%), and a 0\% probability that Kenya achieved the goal.

To examine the performance of  the space-time smoothing model, we held out some of the data and then predicted the U5MR at these points using the weighted estimate and the smoothed estimate. Specifically, we get estimates of the U5MR for all counties and periods from the model using all the 2003 and 2008--2009 DHS, along with 397 clusters from the 2014 DHS. We then calculate weighted estimates of U5MR using the remaining  1,187 clusters, and these are treated as the closest to the truth, since they are based on a large sample. Due to stability of the weighted estimates we look only at the periods  1990--1994,1995--1999, 2000--2004, 2005--2009 and 2010--2014, and form estimates for each of the 47 counties.  Let $Y_{ip}^{(1)}$ denote the weighted estimator and  $Y_{ip}^{(2)}$ the smoothed estimator (from our model) in county $i$ and  period $p$. We compare these estimates with the weighted estimates from the 1,187 clusters, $y_{ip}$.
In particular, we calculate,
\begin{equation}\label{eq:MSE3}
\mbox{MSE}^{(j)}_{p} = \frac{1}{47} \sum_{i=1}^{47
} \left(
Y_{ip}^{(j)} - y_{ip}
\right)^2,
\end{equation}
for $p =\{ $1990--1994,1995--1999, 2000--2004, 2005--2009 $\}$ and $j=1,2$. Table 
\ref{tab:3mse} presents the MSEs and we see that in all cases the smoothing model has far superior performance.

\begin{table}[htp]
\begin{center}
\begin{tabular}{l|l|l}
Period & Weighted & Smoothed\\ \hline
1990--1994 & 55 & 34\\
1995--1999 & 36 & 13\\
2000--2004 & 22 & 6.8\\
2005--2009 & 9.1 & 3.9 \\
2009--2014 & 7.4 & 2.8
\end{tabular}
\end{center}
\caption{Mean-squared errors $(\times 10^4)$ comparing weighted and smoothed estimates, via (\ref{eq:MSE3}).}\label{tab:3mse}
\end{table}%


\section{Exploratory Covariate Modeling}\label{sec:exploratory}

\subsection{The Covariate Model}

In this section we carry out an exploratory investigation into whether any of the spatial variability we see in Kenya can be attributed to a variety of covariates. Before outlining our approach, we provide a brief literature review of  suggestions for building covariate models in the setting considered here. 

\cite{gething:etal:15} describe the use of DHS data to construct surfaces of: access to HIV testing in women, stunting in children, anemia prevalence in children and access to improved sanitation. For each outcome and each country the following procedure was carried out. A collection of 17 covariates  were examined. Initially, simple linear regression was used with three versions  (the original, the square and the square root) of each of the 17 variables taken. Cross-validation was then used to reduce these to a subset of 17 terms.  Two-way interactions for these 17 were added to the collection to give $289=17\times 17$ additional terms. This complete set was reduced  to 20, again via cross-validation. Then the resultant potential $2^{20}-1$ models, that were combinations of these 20 terms, were compared.

\cite{bhatt:etal:17} use an approach known as stacked generalization \citep{wolpert:92} in which multiple predicting algorithms are weighted to produce a final prediction. This approach is closely related to the more general super-learner approach \citep{vanderlaan:07}. This approach is interesting and has optimality properties for prediction but has a lack of interpretability,  the model is not suitable for predictions into the future, and there are questions over whether uncertainty in the procedure can be incorporated into interval estimates for the surface. A similar approach was used by \cite{golding:etal:17}.

\begin{figure}
\centering
\includegraphics[width=\linewidth]{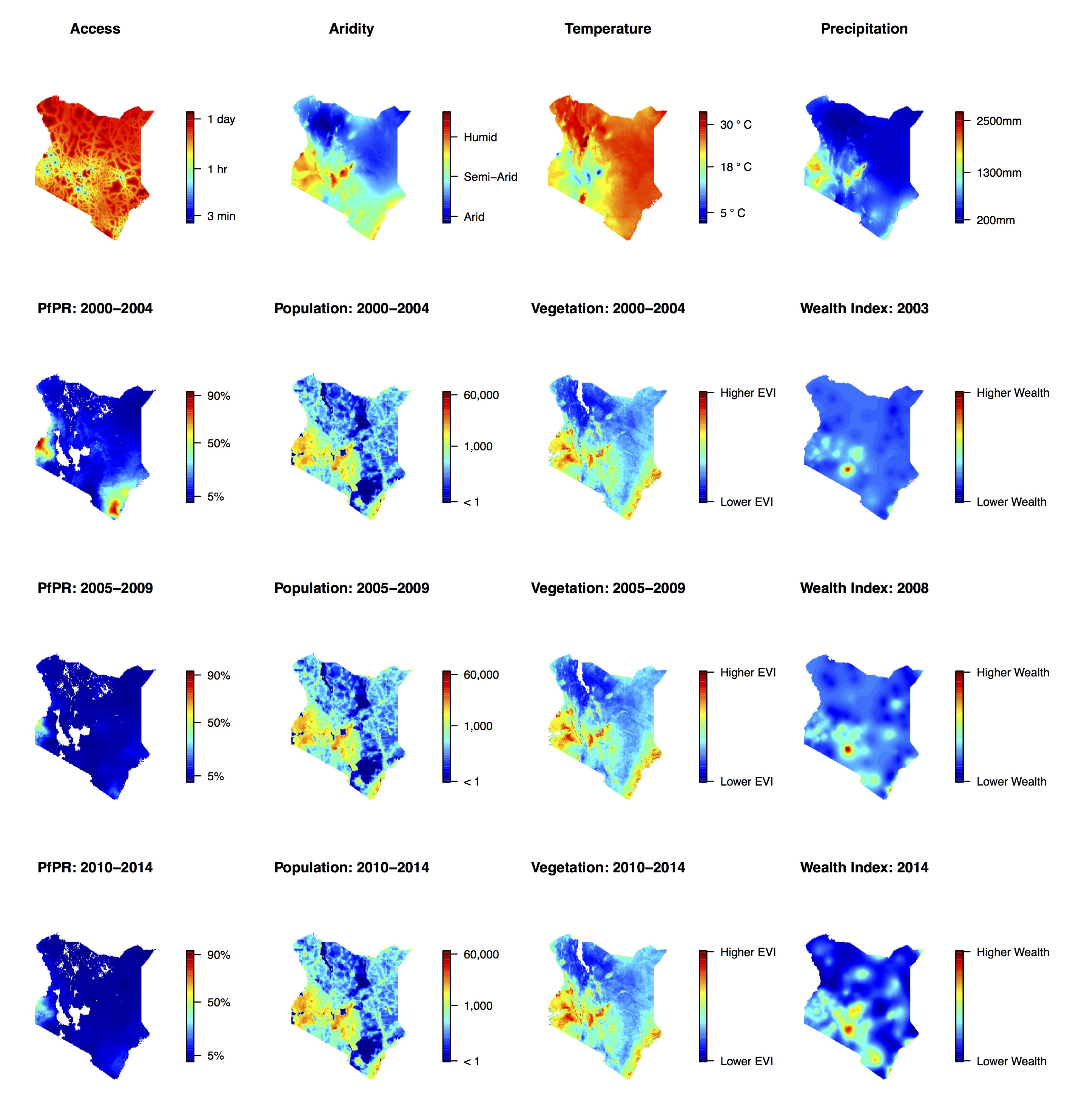}
\caption{Top row: plots of proposed time-invariant spatial covariates in Kenya. Rows 2--4: plots of the average values of proposed time-variant spatial covariates in Kenya over the 3 periods. Access, Aridity, and Population have been log-transformed for presentation purposes. The units for population are number of people per 5km $\times$ 5km area. All time points are on the same scale for each variable.}
\label{Fig:proposed_covs_all}
\end{figure}

The covariates we choose to examine are access \cite[estimated travel time to cities with at least 50,000 people;][]{nelson:2008}, aridity \citep{zomer:2007,zomer:2008}, 
precipitation \citep{fick:2017}, temperature \citep{fick:2017}, enhanced vegetation index \cite[EVI;][]{evidat}, \textit{Plasmodium falciparum} parasite rate (\textit{Pf}PR) in children \citep{bhatt:etal:15}, population \citep{lloyd:2017} and a wealth index (calculated from DHS household data for each and then spatially modeled). Further details on these covariates can be found in the Supplementary Materials.
For the purposes of exploration, we model access, aridity, 
temperature, and precipitation as time-invariant; plots of these variables can be found in
the top row of Figure \ref{Fig:proposed_covs_all}. Data on \textit{Pf}PR, population, and vegetation were obtained for the years 2000--2014 and subsequently averaged within each of the three 5-year periods (2000--2004, 2005--2009 and 2010--2014) to obtain values for each period; these data are also displayed Figure \ref{Fig:proposed_covs_all}.

%

In order to determine which covariates are predictive of U5MR, we will use a simplified version of the model described in Section \ref{sec:reconstruct:model}, in which we replace the yearly model with a model over 5-year periods $p=\{$2000--2004, 2005--2009, 2010--2014$\}$. The model is,
\begin{equation}\label{eq:covmod}
\beta_{a[m],k}(\bms_j,p) = \beta_{a[m]}+\delta_{\text{str}[\scriptsize{\bms_j}]}
 +   \gamma_p +  \eta_j + \upsilon_k +\text{Other Variables}, 
 \end{equation}
where $\beta_{a[m]}$ are the age-specific intercepts, $\delta_{\text{str}[\small{\bms_j}]}$ are stratum (fixed) effects,
$\gamma_p$ is a temporal random effect (assumed common to all age groups) and is modeled using a RW1 (since we have three periods only), $\eta_j \sim_{iid} \mbox{N}(0,\sigma_\eta^2)$ are cluster random effects and $\upsilon_k \sim_{iid} \mbox{N}(0,\sigma_\upsilon^2) $ are survey random effects. In comparisons to be presented in Section \ref{sec:covresults} we compare four different approaches/models with $M_1$ referring to the direct estimates and $M_2,$ $M_3$, $M_4$ corresponding to choosing the ``Other Variables" in (\ref{eq:covmod}) to be space only, covariates only and space and covariates, i.e.,
\begin{align}
S(\bms_j) \label{eq:M2}\tag*{$M_2$}\\
\bmbeta \bx(\bms_j,p)\label{eq:M3}\tag*{$M_3$}\\
\bmbeta \bx (\bms_j,p)+ S(\bms_j)\label{eq:M4}\tag*{$M_4$}
\end{align}
where $\bmx(\bms_j,p)$ are the spatial covariates at location $\bms$ and in period $p$, and $S(\bms_j)$ is a spatial random effect at cluster with location $\bms_j$. The spatial model is as before, a Gaussian Markov random field with M\'atern covariance function (fitted using the SPDE approach) and, for simplicity, we assume it has the same structure for every age group and period.
We divide the data into training and test sets. In the training set we build the models and in the test set we compare their performance. We split the 2014 DHS into two, roughly equal-sized, groups. We use 799 clusters from the 2014 DHS as our test set (for comparison purposes). The other clusters in the 2014 DHS along with data from the 2003 and 2008/2009 DHS will be used for training the model, resulting in 1,581 clusters being used. To emphasize, the spatial model, $M_2$, is fit just once, while $M_3$ and $M_4$ are fit multiple times, for each combination of covariates.
For these models, we assess their performance by using the DIC \citep{spiegelhalter:etal:94}, CPO  \citep{held:etal:10}, and WAIC \citep{watanabe:13} criteria. As a result exercise, we determine the best models in each of the $M_3$ and $M_4$ collections to be used to compare with the direct estimates $M_1$ and spatial model $M_2$ obtained from the training clusters. 
We will have a total of four final comparisons (with all estimates bases on the training data): $M_1$ direct estimates, $M_2$ a model with a spatial random effect, $M_3$  a model with the ``best" collection of covariates, and $M_4$ a model with  alternative ``best" collection of covariates that are chosen when spatial effects are included in the model.

Under $M_j$ we have an estimator of the U5MR for each area $i$ and period $p$, $Y_{ip}^{(j)}$. Under model $M_1$, the direct estimator has normal distribution $\mbox{N}(\widehat{Y}_{ip}^{(1)},V_{ip}^{(1)})$, and under $M_2,M_3,M_4$, we have posterior distributions with posterior means $\widehat{Y}_{ip}^{(j)}$ and posterior variances $V_{ip}^{(j)}$, $j=2,3,4$. Then, with the ``truth" (direct estimate from test data) $y_{ip}$,
$$\mbox{MSE}^{(j)}_{ip} = \mbox{E}\left[ \left(
Y_{ip}^{(j)} - y_{ip}
\right)^2
\right]= \mbox{E}[ \widehat{Y}_{ip}^{(j)}-y_{ip}]^2 + \mbox{var}({Y}_{ip}^{(j)}).
$$
The best approach is that which minimizes the MSE.

\subsection{Exploratory Covariate Modeling Results}\label{sec:covresults}

The DIC, CPO and WAIC scores for all possible covariate combinations for models $M_3$ and $M_4$  are reproduced  in
the Supplementary Materials.
There is good agreement between  the three different assessments of model fit. For $M_3$ (no spatial effects and covariates), the best model was that which included temperature and {\it Pf}PR. For $M_4$ (spatial effects and covariates), the model that included only precipitation and {\it Pf}PR  performed best.

\begin{figure}
\centering
\includegraphics[width=\linewidth]{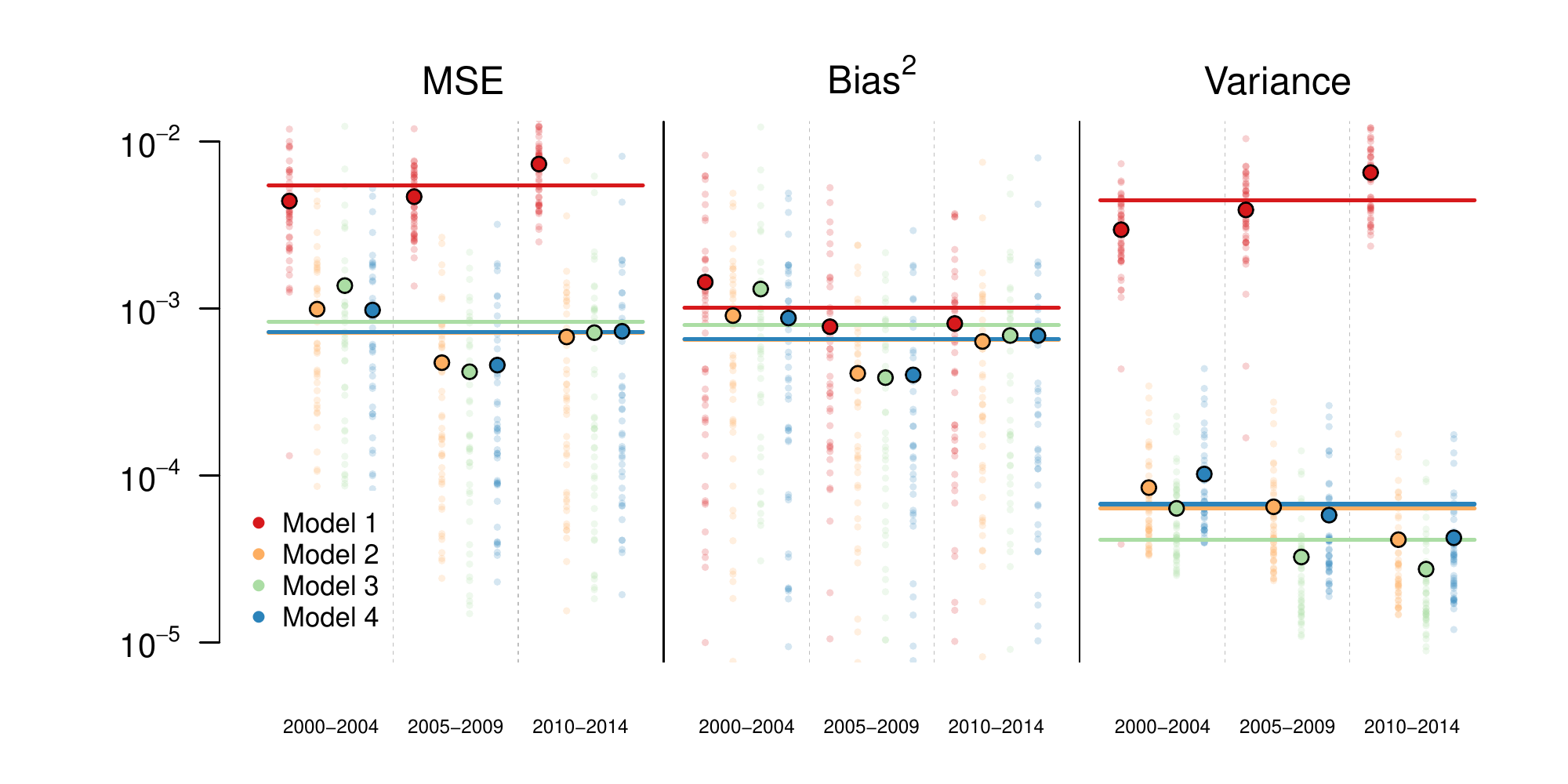}
\caption{Plot of MSE broken down into $\text{Bias}^2$ and variance terms for U5MR. Color coded by model. Horizontal lines indicate the value average over all years. Larger, darker points indicate the average of the 47 admin regions. Note, the y-axis has been transformed and truncated so not all individual values are shown.}\label{fig:covmse}
\end{figure}

\begin{figure}
\centering
\includegraphics[width=0.7\linewidth]{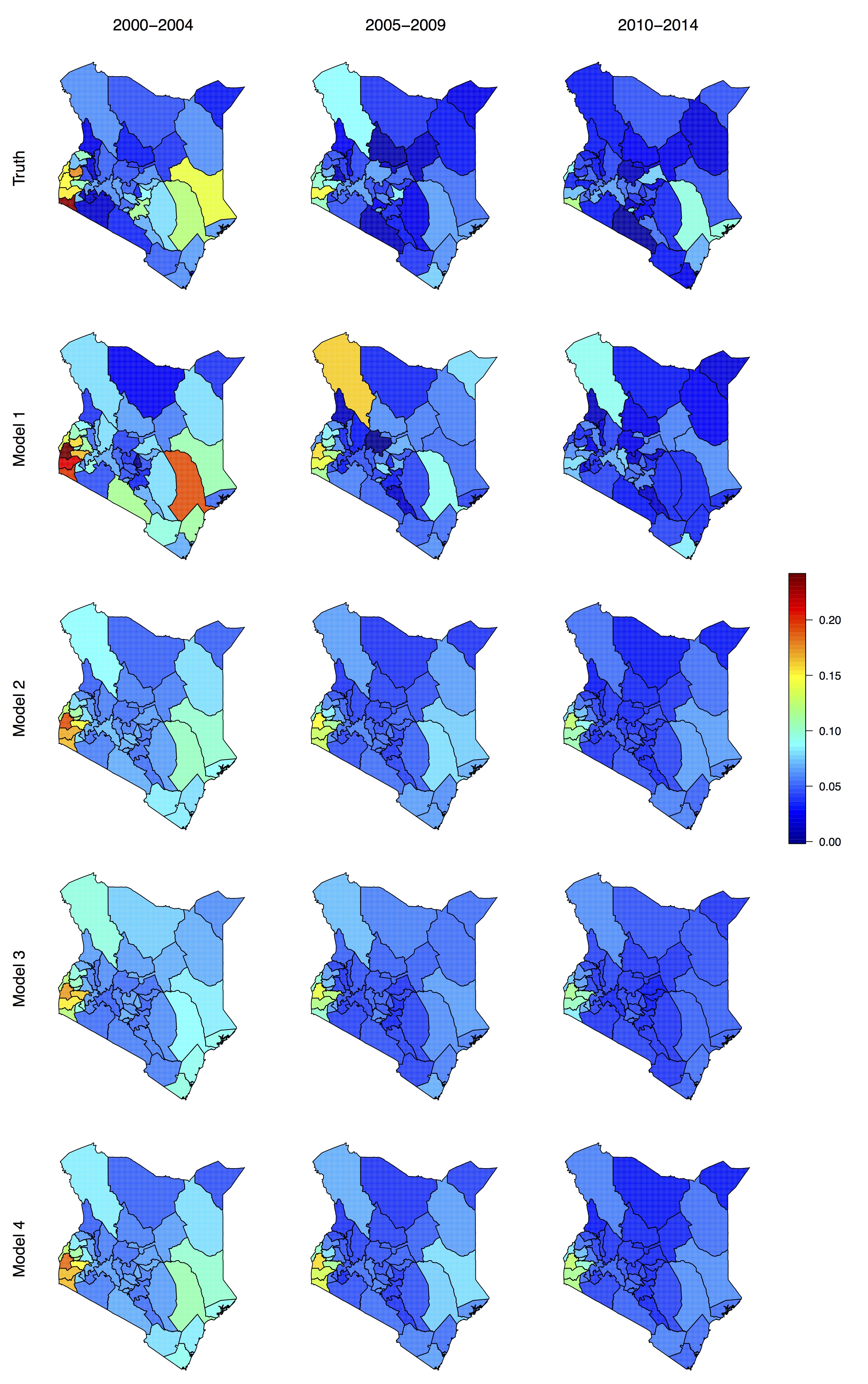}
\caption{Regional predicted U5MR. Top row is the ``truth'', i.e., direct estimates based on the 799 test locations in the 2014 survey. Model $M_1$ are the direct estimates based on the other clusters. Model $M_2$ is the spatial only model (no covariates). Model $M_3$ is the covariates only model (temperature and  {\it Pf}PR). Model $M_4$ is the spatial and covariates model (precipitation and  {\it Pf}PR).} 
\label{fig:aggregate_all}
\end{figure}

The MSE and constituent squared bias and variance are shown in Figure \ref{fig:covmse}. We see that $M_2$, $M_3$ and $M_4$ perform much better than $M_1$, with $M_2$ and $M_4$ being better than $M_3$. We see in Figure \ref{fig:aggregate_all} that the predicted surfaces are almost identical under models $M_3$ and $M_4$. Somewhat surprisingly, the spatial standard deviation and range parameters did not change with the addition of covariates, and the strength of the associations changed little also (see Supplementary Materials). As expected, there is a strong positive association between the logit of U5MR and log {\it Pf}PR. In model $M_3$, the logit of U5MR  also showed a positive association with temperature while in model $M_4$   precipitation was negatively associated.

%

\section{Discussion}\label{sec:discussion}

In this paper we have developed a continuous space/discrete time model for investigating the dynamics of U5MR in a developing world setting. We have illustrated that the model improves on the use of weighted estimates, and can provide reliable inference at the required geographical scale. However, there are a number of aspects that we aim to improve upon in future work. An adjustment for HIV epidemics is crucial, given the extent of the epidemic in Kenya (and in many other countries), and we would like to acknowledge the uncertainty in the bias correction. 

The age pattern of human mortality between ages 0 and 5 years follows a regular, decreasing pattern across a wide range of overall levels.  Net of level, this age pattern can be characterized by the ratio of mortality at each age compared to a reference age.  Our models estimate mortality in six independent age groups, and it is possible that the age pattern that results from combining the estimates from the six models does not follow any of the regularly observed age patterns of human child mortality.  In our analysis, this was not a problem (see Supplementary Materials), but we are currently working on a flexible model of the age pattern of mortality that can enforce this constraint. 

We would like to include other data sources, for both Kenya and other countries. Early DHS do not contain the GPS coordinates of the sampled clusters, but rather the administrative areas within which sampling took place. We plan to extend methods presented in \cite{wilson:wakefield:17} to model the location of the unknown sampling point. WAs described in Section \ref{sec:introduction}, we have utilized so-called full birth history data in which the births and deaths of each child are available. Summary birth history consist of only the number of children ever born and the number who died, by age of mother. These data are easier to collect and are available in a large number of surveys and censuses. The incorporation of such data into a model-based framework is a priority for future work.

In this work we have used a continuous spatial model, whereas our major interest was to inspect results on the discrete scale for the 47 administrative regions. For this purpose we integrated over the spatial field and included the population density to produce the results at the county level. An obvious  question that arises is: what advantages are there with this approach as compared to using a discrete spatial model, such as the ICAR model \citep{besag:etal:91}, directly? One advantage of the continuous model is that we get a smoothed estimated field giving an indication of the U5MR at a finer resolution. Furthermore, the adjustment for the survey design is implicitly integrated into the model when taking the population density into account. It is important to note that this would not be possible using a discrete spatial model.
Another advantage is that when using a continuous random field we do not need to specify a neighborhood structure. The 47 administrative regions of Kenya vary widely in shape and size, and therefore in the number of neighbors, so that it is not clear how to define a sensible neighborhood dependence structure. Part of our future research will be to investigate how a discrete spatial model would perform in this setting. Here, we are particularly interested in the performance of the recently proposed model by \cite{riebler:etal:16} and in the comparison of the results to the continuous model presented here. 

There are several limitations to the covariate modeling carried out in Section \ref{sec:exploratory}. For one, we use geographically-referenced covariates rather than household or individual level variables since we were interested in understanding large-scale patterns in U5MR. Therefore, we do not directly model several variables that are known to have an impact on childhood mortality such as biological factors (e.g.,~vaccination rates, disease prevalence), maternal demographics (e.g.,~age, education), and household characteristics (e.g.,~toilet facilities, access to water). Though spatial surfaces do exist for some of these variables \cite[e.g., measles vaccination coverage:][]{takahashi:2017} or surfaces could be developed based on DHS data \citep{gething:etal:15}, there is greater uncertainty associated with these variables, which can lead to misleading inference \citep{foster:etal:12}. We therefore limited the number of heavily-modeled covariates in our model. Additionally, many of these factors are associated with variables already included in our model.

The computations were run on a computing server with 32 Intel Xeon 2.7 GHz CPUs available. The full Bayesian model required around 14 hours for estimation and 19.5 hours for predictions. An empirical Bayes version of the model required around 2.5 hours for estimation and 10 hours for predictions.
Supplementary materials and code to run the models described here can be found at \url{http://faculty.washington.edu/jonno/software.html}.
 
\section*{Acknowledgments}
Wakefield and Wilson were supported by grant R01CA095994 from the National Institutes of Health, Fuglstad and Riebler  by project number 240873/F20 from the Research Council of Norway, Godwin by R01AI029168 from the National Institutes of Health and Clark by R01HD086227 from the National Institutes of Health. We would like to thank Zehang Li, Yuan Hsiao, Bryan Martin, Danzhen Yu, Lucia Hug, Leontine Alkema, Jon Pedersen, Patrick Gerland, Trevor Croft, Bruno Masquelier, Kenneth Hill, David Sharrow, Roy Burstein, Simon Hay  and Jonathan Muir for providing data and helpful comments.

\clearpage
\bibliographystyle{chicago}
\bibliography{spatepi.bib}

\end{document}

%% file: definitions.tex
\def \s2{\sigma^2}
\def \hs2{\hat{\sigma}^2}
\def \siga2{\sigma_{\alpha}^2}
\def \sige2{\sigma_{\epsilon}^2}
\def \sig2{\sigma^2}

\newcommand{\bms}{\mbox{\boldmath $s$}}

\newcommand{\bmx}{\mbox{\boldmath $x$}}
\newcommand{\bx}{\mbox{\boldmath $x$}}

\newcommand{\beq}{\begin{equation}}       
\newcommand{\eeq}{\end{equation}}       
\newcommand{\beqn}{\begin{eqnarray}}
\newcommand{\eeqn}{\end{eqnarray}}

\newcommand{\bmbeta}{\mbox{\boldmath $\beta$}}

%% file: estimating_under5_mortality.bbl
\begin{thebibliography}{}

\bibitem[\protect\citeauthoryear{Alkema and New}{Alkema and
  New}{2014}]{alkema:new:14}
Alkema, L. and J.~New (2014).
\newblock Global estimation of child mortality using a {B}ayesian {B}-spline
  bias-reduction model.
\newblock {\em The Annals of Applied Statistics\/}~{\em 8}, 2122--2149.

\bibitem[\protect\citeauthoryear{Alkema, New, Pedersen, You, et~al.}{Alkema
  et~al.}{2014}]{alkema2014child}
Alkema, L., J.~R. New, J.~Pedersen, D.~You, et~al. (2014).
\newblock Child mortality estimation 2013: an overview of updates in estimation
  methods by the {U}nited {N}ations {I}nter-{A}gency {G}roup for {C}hild
  {M}ortality {E}stimation.
\newblock {\em PloS ONE\/}~{\em 9}, e101112.

\bibitem[\protect\citeauthoryear{Allison}{Allison}{2014}]{allison:14}
Allison, P. (2014).
\newblock {\em Event History and Survival Analysis, Second Edition}, Volume~46.
\newblock SAGE publications.

\bibitem[\protect\citeauthoryear{Besag, York, and Molli\'e}{Besag
  et~al.}{1991}]{besag:etal:91}
Besag, J., J.~York, and A.~Molli\'e (1991).
\newblock Bayesian image restoration with two applications in spatial
  statistics.
\newblock {\em Annals of the Institute of Statistics and Mathematics\/}~{\em
  43}, 1--59.

\bibitem[\protect\citeauthoryear{Bhatt, Cameron, Flaxman, Weiss, Smith, and
  Gething}{Bhatt et~al.}{2017}]{bhatt:etal:17}
Bhatt, S., E.~Cameron, S.~Flaxman, D.~Weiss, D.~Smith, and P.~Gething (2017).
\newblock Improved prediction accuracy for disease risk mapping using
  {G}aussian process stacked generalization.
\newblock {\em Journal of The Royal Society Interface\/}~{\em 14}, 20170520.

\bibitem[\protect\citeauthoryear{Bhatt, Weiss, Cameron, Bisanzio, Mappin,
  Dalrymple, Battle, Moyes, Henry, Eckhoff, et~al.}{Bhatt
  et~al.}{2015}]{bhatt:etal:15}
Bhatt, S., D.~Weiss, E.~Cameron, D.~Bisanzio, B.~Mappin, U.~Dalrymple,
  K.~Battle, C.~Moyes, A.~Henry, P.~Eckhoff, et~al. (2015).
\newblock The effect of malaria control on plasmodium falciparum in {A}frica
  between 2000 and 2015.
\newblock {\em Nature\/}~{\em 526}, 207--211.

\bibitem[\protect\citeauthoryear{Blangiardo and Cameletti}{Blangiardo and
  Cameletti}{2015}]{blangiardo:cameletti:15}
Blangiardo, M. and M.~Cameletti (2015).
\newblock {\em Spatial and spatio-temporal Bayesian models with R-INLA}.
\newblock John Wiley and Sons.

\bibitem[\protect\citeauthoryear{Burke, Heft-Neal, and Bendavid}{Burke
  et~al.}{2016}]{burke:etal:16}
Burke, M., S.~Heft-Neal, and E.~Bendavid (2016).
\newblock Sources of variation in under-5 mortality across sub-{S}aharan
  {A}frica: a spatial analysis.
\newblock {\em The Lancet Global Health\/}~{\em 4}, e936--e945.

\bibitem[\protect\citeauthoryear{Didan}{Didan}{2015}]{evidat}
Didan, K. (2015).
\newblock {MOD13A3 MODIS/Terra vegetation Indices Monthly L3 Global 1km SIN
  Grid V006}.
\newblock {NASA EOSDIS Land Processes DAAC}.
  \url{https://doi.org/10.5067/modis/mod13a3.006}.

\bibitem[\protect\citeauthoryear{Dwyer-Lindgren, Kakungu, Hangoma, Ng, Wang,
  Flaxman, Masiye, and Gakidou}{Dwyer-Lindgren et~al.}{2014}]{dwyer:etal:2014}
Dwyer-Lindgren, L., F.~Kakungu, P.~Hangoma, M.~Ng, H.~Wang, A.~D. Flaxman,
  F.~Masiye, and E.~Gakidou (2014).
\newblock Estimation of district-level under-5 mortality in {Z}ambia using
  birth history data, 1980--2010.
\newblock {\em Spatial and Spatio-Temporal Epidemiology\/}~{\em 11}, 89--107.

\bibitem[\protect\citeauthoryear{Fick and Hijmans}{Fick and
  Hijmans}{2017}]{fick:2017}
Fick, S.~E. and R.~J. Hijmans (2017).
\newblock {WorldClim} 2: new 1-km spatial resolution climate surfaces for
  global land areas.
\newblock {\em International Journal of Climatology\/}~{\em 37}, 4302--4315.

\bibitem[\protect\citeauthoryear{Foster, Shimadzu, and Darnell}{Foster
  et~al.}{2012}]{foster:etal:12}
Foster, S., H.~Shimadzu, and R.~Darnell (2012).
\newblock Uncertainty in spatially predicted covariates: is it ignorable?
\newblock {\em Journal of the Royal Statistical Society: Series C\/}~{\em 61},
  637--652.

\bibitem[\protect\citeauthoryear{Fuglstad, Simpson, Lindgren, and Rue}{Fuglstad
  et~al.}{2015}]{fuglstad:etal:15}
Fuglstad, G.-A., D.~Simpson, F.~Lindgren, and H.~Rue (2015).
\newblock Constructing priors that penalize the complexity of {G}aussian random
  fields.
\newblock {\em arXiv preprint arXiv:1503.00256\/}.

\bibitem[\protect\citeauthoryear{{{GBD} 2016 {M}ortality
  {C}ollaborators}}{{{GBD} 2016 {M}ortality
  {C}ollaborators}}{2017}]{GBD:mortality:17}
{{GBD} 2016 {M}ortality {C}ollaborators} (2017).
\newblock Global, regional, and national under-5 mortality, adult mortality,
  age-specific mortality, and life expectancy, 1970--2016: a systematic
  analysis for the {G}lobal {B}urden of {D}isease {S}tudy 2016.
\newblock {\em The Lancet\/}~{\em 390}, 1084--1150.

\bibitem[\protect\citeauthoryear{Gething, Tatem, Bird, and
  Burgert-Brucker}{Gething et~al.}{2015}]{gething:etal:15}
Gething, P., A.~Tatem, T.~Bird, and C.~Burgert-Brucker (2015).
\newblock Creating spatial interpolation surfaces with {DHS} data.
\newblock Technical report, ICF International.
\newblock DHS Spatial Analysis Reports No. 11.

\bibitem[\protect\citeauthoryear{Golding, Burstein, Longbottom, Browne,
  Fullman, Osgood-Zimmerman, Earl, Bhatt, Cameron, Casey, Dwyer-Lindgren,
  Farag, Flaxman, Fraser, Gething, Gibson, Graetz, Krause, Kulikoff, Lim,
  Mappin, Morozoff, Reiner, Sligar, Smith, Wang, Weiss, Murray, Moyes, and
  Hay}{Golding et~al.}{2017}]{golding:etal:17}
Golding, N., R.~Burstein, J.~Longbottom, A.~Browne, N.~Fullman,
  A.~Osgood-Zimmerman, L.~Earl, S.~Bhatt, E.~Cameron, D.~Casey,
  L.~Dwyer-Lindgren, T.~Farag, A.~Flaxman, M.~Fraser, P.~Gething, H.~Gibson,
  N.~Graetz, L.~Krause, X.~Kulikoff, S.~Lim, B.~Mappin, C.~Morozoff, R.~Reiner,
  A.~Sligar, D.~Smith, H.~Wang, D.~Weiss, C.~Murray, C.~Moyes, and S.~Hay
  (2017).
\newblock Mapping under-5 and neonatal mortality in {A}frica, 2000--15: a
  baseline analysis for the {S}ustainable {D}evelopment {G}oals.
\newblock {\em The Lancet\/}.
\newblock Available online, September 25th, 2017.

\bibitem[\protect\citeauthoryear{Haakenstad, Birger, Singh, Liu, Lim, Ng, and
  Dieleman}{Haakenstad et~al.}{2016}]{haakenstad:etal:16}
Haakenstad, A., M.~Birger, L.~Singh, P.~Liu, S.~Lim, M.~Ng, and J.~Dieleman
  (2016).
\newblock Vaccine assistance to low-and middle-income countries increased to
  \$3.6 billion in 2014.
\newblock {\em Health Affairs\/}~{\em 35}, 242--249.

\bibitem[\protect\citeauthoryear{Hallett, Anderson, Asante, Bartlett, Bendaud,
  Bhatt, Burgert, Cuadros, Dzangare, Fecht, et~al.}{Hallett
  et~al.}{2016}]{hallett:etal:16}
Hallett, T., S.-J. Anderson, C.~A. Asante, N.~Bartlett, V.~Bendaud, S.~Bhatt,
  C.~Burgert, D.~F. Cuadros, J.~Dzangare, D.~Fecht, et~al. (2016).
\newblock Evaluation of geospatial methods to generate subnational {HIV}
  prevalence estimates for local level planning.
\newblock {\em AIDS\/}~{\em 30}, 1467--1474.

\bibitem[\protect\citeauthoryear{Held, Schr{\"o}dle, and Rue}{Held
  et~al.}{2010}]{held:etal:10}
Held, L., B.~Schr{\"o}dle, and H.~Rue (2010).
\newblock Posterior and cross-validatory predictive checks: A comparison of
  {MCMC} and {INLA}.
\newblock In T.~Kneib and G.~Tutz (Eds.), {\em Statistical Modeling and
  Regression Structures -- Festschrift in Honour of Ludwig Fahrmeir}, pp.\
  91--110. Physica-Verlag.

\bibitem[\protect\citeauthoryear{Horvitz and Thompson}{Horvitz and
  Thompson}{1952}]{horvitz:thompson:52}
Horvitz, D. and D.~Thompson (1952).
\newblock A generalization of sampling without replacement from a finite
  universe.
\newblock {\em Journal of the American Statistical Association\/}~{\em 47},
  663--685.

\bibitem[\protect\citeauthoryear{Knorr-Held}{Knorr-Held}{2000}]{knorrheld:00}
Knorr-Held, L. (2000).
\newblock Bayesian modelling of inseparable space-time variation in disease
  risk.
\newblock {\em Statistics in Medicine\/}~{\em 19}, 2555--2567.

\bibitem[\protect\citeauthoryear{Larmarange and Bendaud}{Larmarange and
  Bendaud}{2014}]{larmarange:bendaud:14}
Larmarange, J. and V.~Bendaud (2014).
\newblock {HIV} estimates at second subnational level from national
  population-based surveys.
\newblock {\em AIDS\/}~{\em 28}, S469--S476.

\bibitem[\protect\citeauthoryear{Leroux, Lei, and Breslow}{Leroux
  et~al.}{1999}]{leroux:etal:99}
Leroux, B., X.~Lei, and N.~Breslow (1999).
\newblock Estimation of disease rates in small areas: A new mixed model for
  spatial dependence.
\newblock In M.~Halloran and D.~Berry (Eds.), {\em Statistical Models in
  Epidemiology, the Environment and Clinical Trials}, pp.\  179--192. New York:
  Springer.

\bibitem[\protect\citeauthoryear{Linard, Gilbert, Snow, Noor, and Tatem}{Linard
  et~al.}{2012}]{linard:etal:12}
Linard, C., M.~Gilbert, R.~W. Snow, A.~M. Noor, and A.~J. Tatem (2012).
\newblock Population distribution, settlement patterns and accessibility across
  africa in 2010.
\newblock {\em PloS One\/}~{\em 7}, e31743.

\bibitem[\protect\citeauthoryear{Lindgren, Rue, and Linstr\"{o}m}{Lindgren
  et~al.}{2011}]{lindgren:etal:11}
Lindgren, F., H.~Rue, and J.~Linstr\"{o}m (2011).
\newblock An explicit link between {G}aussian fields and {G}aussian {M}arkov
  random fields: the stochastic differential equation approach (with
  discussion).
\newblock {\em Journal of the Royal Statistical Society, Series B\/}~{\em 73},
  423--498.

\bibitem[\protect\citeauthoryear{Lloyd, Sorichetta, and Tatem}{Lloyd
  et~al.}{2017}]{lloyd:2017}
Lloyd, C., A.~Sorichetta, and A.~Tatem (2017).
\newblock High resolution global gridded data for use in population studies.
\newblock {\em Scientific Data\/}~{\em 4}, 170001.

\bibitem[\protect\citeauthoryear{Mercer, Wakefield, Pantazis, Lutambi, Mosanja,
  and Clark}{Mercer et~al.}{2015}]{mercer:etal:15}
Mercer, L., J.~Wakefield, A.~Pantazis, A.~Lutambi, H.~Mosanja, and S.~Clark
  (2015).
\newblock Small area estimation of childhood of childhood mortality in the
  absence of vital registration.
\newblock {\em Annals of Applied Statistics\/}~{\em 9}, 1889--1905.

\bibitem[\protect\citeauthoryear{Nelson}{Nelson}{2008}]{nelson:2008}
Nelson, A. (2008).
\newblock Estimated travel time to the nearest city of 50,000 or more people in
  year 2000.
\newblock Technical report, Global Environment Monitoring Unit - Joint Research
  Centre of the European Commission, Ispra, Italy.

\bibitem[\protect\citeauthoryear{Pezzulo, Utazi, Sorichetta, Tatem,
  Yourkavitch, Pullum, and Burgert-Brucker}{Pezzulo
  et~al.}{2017}]{pezzulo:etal:17}
Pezzulo, C., E.~Utazi, T.~B.~A. Sorichetta, A.~Tatem, J.~Yourkavitch,
  T.~Pullum, and C.~Burgert-Brucker (2017).
\newblock Subnational modelling of child mortality and its drivers across 27
  countries in {S}ub-{S}aharan {A}frica.
\newblock Technical report, Paper presented at PAA Meeting.

\bibitem[\protect\citeauthoryear{Rao and Molina}{Rao and
  Molina}{2015}]{rao:molina:15}
Rao, J. and I.~Molina (2015).
\newblock {\em Small Area Estimation, Second Edition}.
\newblock New York: John Wiley.

\bibitem[\protect\citeauthoryear{Riebler, S{\o}rbye, Simpson, and Rue}{Riebler
  et~al.}{2016}]{riebler:etal:16}
Riebler, A., S.~H. S{\o}rbye, D.~Simpson, and H.~Rue (2016).
\newblock An intuitive {B}ayesian spatial model for disease mapping that
  accounts for scaling.
\newblock {\em Statistical Methods in Medical Research\/}~{\em 25\/}(4),
  1145--1165.

\bibitem[\protect\citeauthoryear{Rue, Martino, and Chopin}{Rue
  et~al.}{2009}]{rue:etal:09}
Rue, H., S.~Martino, and N.~Chopin (2009).
\newblock Approximate {B}ayesian inference for latent {G}aussian models using
  integrated nested {L}aplace approximations (with discussion).
\newblock {\em Journal of the Royal Statistical Society, Series B\/}~{\em 71},
  319--392.

\bibitem[\protect\citeauthoryear{Simpson, Rue, Riebler, Martins, and
  S{\o}rbye}{Simpson et~al.}{2017}]{simpson:etal:17}
Simpson, D., H.~Rue, A.~Riebler, T.~Martins, and S.~S{\o}rbye (2017).
\newblock Penalising model component complexity: A principled, practical
  approach to constructing priors (with discussion).
\newblock {\em Statistical Science\/}~{\em 32}, 1--28.

\bibitem[\protect\citeauthoryear{Skinner and Wakefield}{Skinner and
  Wakefield}{2017}]{skinner:wakefield:17}
Skinner, C. and J.~Wakefield (2017).
\newblock Introduction to the design and analysis of complex survey data.
\newblock {\em Statistical Science\/}~{\em 32}, 165--175.

\bibitem[\protect\citeauthoryear{Spiegelhalter, Freedman, and
  Parmar}{Spiegelhalter et~al.}{1994}]{spiegelhalter:etal:94}
Spiegelhalter, D., L.~Freedman, and M.~Parmar (1994).
\newblock Bayesian approaches to randomized trials (with discussion).
\newblock {\em Journal of the Royal Statistical Society, Series A\/}~{\em 157},
  357--416.

\bibitem[\protect\citeauthoryear{Takahashi, Metcalf, Ferrari, Tatem, and
  Lessler}{Takahashi et~al.}{2017}]{takahashi:2017}
Takahashi, S., C.~J.~E. Metcalf, M.~J. Ferrari, A.~Tatem, and J.~Lessler
  (2017).
\newblock The geography of measles vaccination in the african great lakes
  region.
\newblock {\em Nature Communications\/}~{\em 8}.

\bibitem[\protect\citeauthoryear{Van~der Laan, Polley, and Hubbard}{Van~der
  Laan et~al.}{2007}]{vanderlaan:07}
Van~der Laan, M.~J., E.~C. Polley, and A.~E. Hubbard (2007).
\newblock Super learner.
\newblock {\em Statistical Applications in Genetics and Molecular
  Biology\/}~{\em 6}.

\bibitem[\protect\citeauthoryear{Wakefield}{Wakefield}{2008}]{wakefield:08}
Wakefield, J. (2008).
\newblock Ecologic studies revisited.
\newblock {\em Annual Review of Public Health\/}~{\em 29}, 75--90.

\bibitem[\protect\citeauthoryear{Wakefield, Simpson, and Godwin}{Wakefield
  et~al.}{2016}]{wakefield:simpson:godwin:16}
Wakefield, J., D.~Simpson, and J.~Godwin (2016).
\newblock {C}omment: Getting into space with a weight problem. {D}iscussion of,
  ``{M}odel-based geostatistics for prevalence mapping in low-resource
  settings", by {P.J. D}iggle and {E. G}iorgi.
\newblock {\em Journal of the American Statistical Association\/}~{\em 111},
  1111--1119.

\bibitem[\protect\citeauthoryear{Walker, Hill, and Zhao}{Walker
  et~al.}{2012}]{walker:etal:12}
Walker, N., K.~Hill, and F.~Zhao (2012).
\newblock Child mortality estimation: methods used to adjust for bias due to
  {AIDS} in estimating trends in under-five mortality.
\newblock {\em PLoS Med\/}~{\em 9}, e1001298.

\bibitem[\protect\citeauthoryear{Watanabe}{Watanabe}{2013}]{watanabe:13}
Watanabe, S. (2013).
\newblock A widely applicable {B}ayesian information criterion.
\newblock {\em Journal of Machine Learning Research\/}~{\em 14}, 867--897.

\bibitem[\protect\citeauthoryear{Wilson and Wakefield}{Wilson and
  Wakefield}{2017}]{wilson:wakefield:17}
Wilson, K. and J.~Wakefield (2017).
\newblock Pointless continuous spatial surface reconstruction.
\newblock {\em ar{X}iv:1709.09659\/}.

\bibitem[\protect\citeauthoryear{Wolpert}{Wolpert}{1992}]{wolpert:92}
Wolpert, D. (1992).
\newblock Stacked generalization.
\newblock {\em Neural Networks\/}~{\em 5}, 241--259.

\bibitem[\protect\citeauthoryear{{W}orld{P}op}{{W}orld{P}op}{2017}]{worldpop:births:17}
{W}orld{P}op (2017).
\newblock Kenya 1km births, version 2.
\newblock Technical report, University of Southampton, DOI:
  10.5258/SOTON/WP00349.

\bibitem[\protect\citeauthoryear{Zomer, Bossio, Trabucco, Yuanjie, Gupta, and
  Singh}{Zomer et~al.}{2007}]{zomer:2007}
Zomer, R.~J., D.~A. Bossio, A.~Trabucco, L.~Yuanjie, D.~C. Gupta, and V.~P.
  Singh (2007).
\newblock Trees and water: smallholder agroforestry on irrigated lands in
  {Northern India}.
\newblock Technical report, International Water Management Institute.

\bibitem[\protect\citeauthoryear{Zomer, Trabucco, Bossio, and Verchot}{Zomer
  et~al.}{2008}]{zomer:2008}
Zomer, R.~J., A.~Trabucco, D.~A. Bossio, and L.~V. Verchot (2008).
\newblock Climate change mitigation: A spatial analysis of global land
  suitability for clean development mechanism afforestation and reforestation.
\newblock {\em Agriculture, Ecosystems and Environment\/}~{\em 126}, 67--80.

\end{thebibliography}
